# Nonvolatile Electric-Field Control of Inversion Symmetry

Lucas Caretta[1,12,*], Yu-Tsun Shao[2,12], Jia Yu[3], Antonio B. Mei[4], Bastien F. Grosso[5], Cheng Dai[6], Piush Behera[1], Daehun Lee[3], Margaret McCarter[7], Eric Parsonnet[7], Harikrishnan K.P.[2], Fei Xue[6], Ed Barnard[8], Steffen Ganschow[9], Archana Raja[8], Lane W. Martin[1,10], Long-Qing Chen[6], Manfred Fiebig[5], Keji Lai[3], Nicola A. Spaldin[5], David A. Muller[2,11], Darrell G. Schlom[4,9,11], Ramamoorthy Ramesh[1,7,10,*]

[1]*Department of Materials Science and Engineering, University of California, Berkeley, California, USA*
[2]*School of Applied and Engineering Physics, Cornell University, Ithaca, New York, USA.*
[3]*Department of Physics, University of Texas, Austin, Texas, USA*
[4]*Department of Materials Science and Engineering, Cornell University, Ithaca, New York, USA.*
[5]*Department of Materials, ETH Zurich, Zurich, Switzerland.*
[6]*Department of Materials Science and Engineering, The Pennsylvania State University, University Park, Pennsylvania, USA*
[7]*Department of Physics, University of California, Berkeley, California, USA*
[8]*Molecular Foundry, Lawrence Berkeley National Laboratory, Berkeley, California, USA*
[9]*Leibniz-Institut für Kristallzüchtung, Max-Born-Str. 2, Berlin, Germany*
[10]*Materials Sciences Division, Lawrence Berkeley National Laboratory, Berkeley, California, USA*
[11]*Kavli Institute at Cornell for Nanoscale Science, Ithaca, New York, USA*
[12]*These authors contributed equally.*

*Correspondence to: rramesh@berkeley.edu, caretta@berkeley.edu

## Abstract

**In condensed-matter systems, competition between ground states at phase boundaries can lead to significant changes in material properties under external stimuli[1–4], particularly when these ground states have different crystal symmetries. A key scientific and technological challenge is to stabilize and control coexistence of symmetry-distinct phases with external stimuli. Using $BiFeO_3$ (BFO) layers confined between layers of the dielectric $TbScO_3$ as a model system, we stabilize the mixed-phase coexistence of centrosymmetric and non-centrosymmetric BFO phases with antipolar, insulating and polar, semiconducting behavior, respectively at room temperature. Application of in-plane electric (polar) fields can both remove and introduce centrosymmetry from the system resulting in reversible, nonvolatile interconversion between the two phases. This interconversion between the centrosymmetric insulating and non-centrosymmetric semiconducting phases coincides with simultaneous changes in the non-linear optical response of over three orders of magnitude, a change in resistivity of over five orders of magnitude, and a change in the polar order. Our work establishes**

**a materials platform allowing for novel cross-functional devices which take advantage of changes in optical, electrical, and ferroic responses.**

## Introduction

Crystal symmetry in condensed-matter materials largely dictates their micro- and macro-scopic properties[5], and in turn their functionalities. Much effort has been devoted to designing and tuning symmetry in solid-state materials[6–12], with ferroelectrics being a particularly pervasive example of the manifestation of broken inversion symmetry[13] (Extended Data Fig. 1). The ability to synthesize heteroepitaxial ferroelectrics has enabled new methods to control ferroic order and even crystal symmetry[14]. Growth of ferroelectric superlattices using layer-by-layer deposition techniques and the utilization of *ab initio* calculations have led to discoveries of nontrivial polar textures[15–19], room-temperature magnetoelectric materials[20], improper ferroelectrics[21–25], antipolar phases[26,27], and correlated phenomena[28]. Such new phases are stabilized by an interplay of strain, electrostatic, and gradient energies as a consequence of the boundary conditions imposed through coherent heteroepitaxy. One can then ask the question: are there pathways to go between polar and antipolar states that contain their own independent crystal symmetries and order parameter-energy landscapes[29,30]? Prior experimental and theoretical work has demonstrated the ability to manipulate inversion symmetry breaking, crystal structures, and polar order in materials with external stimuli, such as electric fields[31,32], resonant optical excitations[11,12,33], and strain[34,35]. However, the ability to both *remove* and *introduce* symmetry with an electric field remains elusive, as electric fields generally only break symmetries.

Here, we show that this interconversion is indeed possible using confined $BiFeO_3$ (BFO) layers as our model system. The heterointerface boundary conditions that we utilize are: (i) a discontinuity of the local polarization at the heterointerface; (ii) a lattice mismatch arising from the different lattice parameters of the two layers; and (iii) a discontinuity in the sense and direction of the octahedral rotations in the two layers, which we achieve by using $TbScO_3$ (TSO) as our epitaxial dielectric interleaving layers. These boundary conditions imposed on the confined BFO layers enable us to stabilize a mixed-phase coexistence of a non-centrosymmetric polar and a centrosymmetric antipolar phase. Both phases are identified and characterized using a combination of high-resolution and four-dimensional (4D) scanning transmission electron microscopy (STEM), piezoresponse force microscopy (PFM), microwave impedance microscopy (MIM), confocal second harmonic generation (SHG), transport measurements, and density-functional theory (DFT). Moreover, using applied electric fields, we interconvert between the polar non-centrosymmetric and antipolar centrosymmetric phases reversibly and in a nonvolatile fashion. With this electric field-driven phase transformation, we show that the applied electric field not only removes centrosymmetry to stabilize polar order, but an electric (*i.e.*, polar) field can remarkably re-introduce centrosymmetry and stabilize antipolar order. We further demonstrate that electric-field manipulation of the crystal symmetry also manifests as concomitant non-volatile, dramatic changes in both the second harmonic generated (SHG) signal and the DC and microwave conductivity. Such large changes in functional materials properties as a result of an electric field-induced symmetry phase transformation open pathways to new opto-electronic devices and highlight a new design scheme and materials platform for developing phase-change-based memory and logic.

**Mixed-Phase Coexistence of Non-centrosymmetric and Centrosymmetric Phases**

Recent studies have shown that various low-energy phases and polytypes of BFO can be stabilized using different experimentally achievable boundary conditions[24,26,27,36–39]. The calculated energies and structures of such phases as a function of their respective unit cell sizes (Methods, Extended data Fig. S21) are shown in Fig. 1a. Of note is the presence of phases with different crystal symmetries in energetic proximity to the *R*3*c* ground state, including a centrosymmetric *Pnma* phase, as well as large unit cell, non-centrosymmetric *Pc* phases with polarization waves in which the relative stability is a function of the wavelength.

The energetic proximity of these BFO phases with different symmetry suggests the possibility of stabilizing multiple coexisting phases with significantly different properties. To investigate this possibility, epitaxial superlattices of $[(BFO)_n/(TSO)_m]_{20}$ for $n = 11 – 20$, $m = 10$ unit cells were synthesized by reactive molecular beam epitaxy (MBE) on $GdScO_3$ and TSO $[110]_O$ substrates (the O denotes orthorhombic orientation; Methods). High-angle annular dark field (HAADF) STEM imaging of a $[(BFO)_{14}/(TSO)_{10}]_{20}$//GSO superlattice shows atomically sharp interfaces, with no obvious presence of crystallographic defects (Fig. 1b, Extended Data Fig. 2, and the X-ray reciprocal space maps in Extended Data Fig. 3). To study the symmetry of the BFO layers, we employed scanning convergent beam electron diffraction (SCBED) coupled with an electron microscopy pixel array detector (EMPAD)[40,41] (Methods). The bright-field (BF) image reconstructed from the SCBED dataset (Fig. 1c) reveals the coexistence of two distinct phases within the BFO layers exhibiting bright and dark contrast, respectively. A symmetry analysis (Extended Data Fig. 4) identifies the two phases as the non-centrosymmetric *Pc* phase (dark regions, Fig. 1c) and a centrosymmetric *Pnma* phase (bright regions, Fig. 1c).

In polar crystals, the charge redistribution associated with ferroelectric polarization leads to intensity asymmetry in polarity-sensitive Friedel pairs of Bragg reflections in the CBED pattern; hence, SCBED in combination with dynamical diffraction simulations enables polarization mapping at sub-nanometer resolution[42–44] (Supplementary Text 1). The arrows map the polarization from the non-centrosymmetric *Pc* BFO phase (Fig. 1d corresponding to the dark regions, Fig. 1c), which shows continuously winding electric dipoles resembling a polarization wave or a series of half vortices. The continuous rotation of the polarization can be represented by $\nabla \times \boldsymbol{P}$ (denoted by the blue/red color overlay, Fig. 1d). In addition to the rotating polarization within the wave, there is a net in-plane polarization along the wave direction ($[100]_{pc}$ or $[\bar{1}00]_{pc}$) that can vary in antiparallel direction between and within each BFO layer in the superlattice. The atomic structure of the coexisting centrosymmetric *Pnma* phase was also probed (Fig. 1e and f, corresponding to the bright regions, Fig. 1c) by HAADF-STEM images along two projections of the crystallographic zone axes (*i.e.*, $[100]_{pc}$ and $[010]_{pc}$, where pc denotes pseudocubic indices). Overlaid on these atomic images are the polar-vector maps of the bismuth-ion displacement, which shows the antipolar "up-up-down-down" order along the $[011]_{pc}$ projection also observed elsewhere[26]. The ferroelectric and antiferroelectric (AFE) nature of the two phases is confirmed with charge-voltage hysteresis loops (Extended Data Fig. 5 and see also Supplementary Text 2). Therefore, we have stabilized two low-energy phases of BFO with non-centrosymmetric polar and centrosymmetric antipolar properties, where the net polarization of the polarization wave ($[100]_{pc}$ or $[\bar{1}00]_{pc}$) phase orients orthogonally to the local dipoles in the antipolar phase ($<011>_{pc}$). HAADF-STEM taken along the $[100]_{pc}$ zone axis reveals the atomically sharp boundary between the coexisting BFO phases (Fig. 1g and Extended Data Fig. 6), suggestive of a first-order phase transformation[45].

Such mixtures of nearly energetically degenerate phases have been the framework used to elicit significant responses to external stimuli[1–4,46]. We now demonstrate such large, simultaneous changes in piezoelectricity, optical SHG, microwave response, and DC conductivity in the mixed-phase superlattices. Lateral PFM phase imaging (Fig. 2a), in which the contrast is sensitive to in-plane polarization along the $[100]_{pc}$ of the uppermost BFO layer in the sample reveals distinct, stripe-like regions of high piezoelectric response (white and dark brown regions) and zero piezoelectric response (orange regions) on the order of several micrometers in width and extending for hundreds micrometers along the $[010]_{pc}$. Consistent with the HR-STEM vector mapping, the high piezoresponse regions are comprised of the polar phase of BFO, where white subdomains have net polarization along the $[\bar{1}00]_{pc}$ and dark-brown subdomains have net polarization along the antiparallel $[100]_{pc}$ (Fig. 2a). Regions of zero piezoresponse (orange regions, Fig. 2a) correspond to the antipolar BFO phase. These observations are confirmed with additional lateral and out-of-plane PFM measurements (Extended Data Fig. 7). We note that the relative stability of each phase can be controlled by tuning the thickness of the BFO layers in the superlattice. As shown by lateral PFM images (Extended Data Fig. 8 and 9) and confirmed with phase field simulations (Extended Data Fig. S23), small unit cell changes in the BFO thickness from 11 to 19 unit cells transform the superlattice from a uniform antipolar state[26] to a uniform polar state, while changing the strain state via substrate selection changes the length scales of the domains (Supplementary Text 3).

**Nonlinear Optical Response of Mixed-Phase Coexistence**

The coexistence of polar and antipolar symmetries also results in a dramatic spatial variation of the non-linear optical response of the material. A SHG map (Methods) of a nearby region on the same sample (Fig. 2b) provides, unlike PFM, information that is integrated throughout the thickness of the film[47]. Furthermore, the comparison of PFM and SHG signals helps to distinguish true polarization effects from band-structural changes of the piezoelectric or nonlinear optical susceptibilities that may occur because of the phase transition. We select a normal-incidence optical geometry, for which the transversally polarized fundamental and, thus, SHG light probes the in-plane symmetry breaking. Distinct centrosymmetric antipolar (dark intensity) and non-centrosymmetric polar (bright intensity) stripe-like phases of BFO are also seen (Fig. 2b, also see Extended Data Fig. S20). The distributions of the brightness within the polar regions differ, however, because of SHG interference occurring at the domain walls and the possibility of domains of different polarization stacked perpendicular to the film surface. Local SHG polar plots from both BFO phases, in which the incident light polarization is varied and the corresponding vertically (Fig. 2c) or horizontally (Fig. 2d) polarized emitted light at the second harmonic is analyzed, are also provided. The non-centrosymmetric polar phase (red circles, Fig. 2c,d) shows a two-lobed angle-dependent SHG intensity profile, with two additional minor lobes, consistent with the space group *Pc* (point-group symmetry *m*) (Fig. 1a) obtained from the CBED analysis (Extended Data Fig. 4 and SHG selection rules in Methods). On the other hand, the centrosymmetric antipolar phase (green squares, Fig. 2c,d) has no measurable SHG signal regardless of the polarization of the incoming light or analyzed SHG light, as expected from the *Pnma* space group (point-group *mmm*) observed in the CBED analysis (Extended Data Fig. 4). Additional PFM phase images and SHG maps are provided (Extended Data Figs. 10 and 11) which scrutinize the spatial correlation of the PFM and SHG signals on the sample. The dramatic difference in non-linear optical response from each phase is further highlighted in the line scan (Fig. 2e, dashed white arrow of Fig. 2b), where the SHG intensity is enhanced by nearly five orders of magnitude in the polar phase relative to the antipolar phase. Consistent with the similar energies of the two phases found in *ab-initio* calculations (Fig. 1a), the piezoelectric, SHG, and CBED data confirm the mixed-phase coexistence of a centrosymmetric antipolar phase and a non-centrosymmetric polar phase.

**Dielectric Response and Conductivity of Mixed-Phase Coexistence**

We explore the changes in dielectric permittivity accompanied by such changes in symmetry, which are often observed during phase transitions in ferroelectric systems (*e.g.*, in a temperature-driven phase transition)[7]. The spatially resolved dielectric response and AC conductivity of the superlattices was probed by scanning microwave impedance microscopy[48] (MIM; Methods). First, by performing PFM with the shielded MIM tip[49], we independently confirmed the coexistence of the mixed polar and antipolar phases in a separate $[(BFO)_{14}/(TSO)_{10}]_{20}$ superlattice grown on a TSO substrate (Fig. 3a). We note that despite the growth of this nominally identical superlattice on a different substrate (strain state), the mixed-phase coexistence is persistent, only manifesting as a difference in length scales of the domains and not any differences in the nature of the phases formed (also see Extended Data Fig. S9). The imaginary (MIM-Im; Fig. 3c) and real (MIM-Re; Fig. 3c) parts of the 2.513 GHz MIM impedance were acquired on the same area as the PFM. The one-to-one correlation between the PFM and MIM images is apparent, with the polar phase showing a significantly enhanced signal compared to the antipolar phase, regardless of the polarization direction. We estimated the dielectric contribution to the MIM-Im contrast by performing finite-element analysis (FEA)[50] of the near-field interaction for the specific tip-sample configuration (Extended Data Fig. 12, and Supplementary Text 4). We plot the simulated MIM-Im signal as a function of the permittivity of the BFO (Extended Data Fig. 12c). Assuming a dielectric constant $\varepsilon_r \sim 60$ in the polar phase[51,52], we estimated the relative dielectric constants between the two phases by comparing FEA simulation with the magnitude of experimentally-observed MIM-Im signal (Fig. 3b), yielding $\varepsilon_r \approx 30$ for the antipolar phase. This is qualitatively consistent with independent dielectric measurements at low frequencies (Extended Data Fig. 13 and Supplementary Text 5). Next, to determine the conductivity contribution to the MIM contrast, we plot the simulated Re- and Im-MIM signals as a function of sample conductivity[50] ($\sigma$; Extended Data Fig. 12d), with the dielectric contribution to the MIM signal (Extended Data Fig. 12c) accounted for by a vertical shift in the MIM-Im curve relative to the MIM-Re curve. Importantly, the small but measurable contrast in the experimentally-observed MIM-Re channel (Fig. 3c) is indicative of a finite GHz conductivity in the non-centrosymmetric polar phase (Supplementary Text 4 and Extended Data Fig. 12). Surprisingly, the simulated $\sigma$-dependent MIM contrast (Extended Data Fig. 12) suggests that the polar phase has a conductivity of ~1 S/m (*i.e.*, a resistivity of ~100 Ω·cm), over the frequency range of 100 MHz to 3 GHz (Extended Data Fig. 12e,f). The width of the grey region denotes the range of σ values within experimental uncertainty. Since this resistivity is approximately 5-6 orders of magnitude lower than that of bulk BFO, the MIM results strongly suggest that the polar phase displays semiconducting behavior, whereas the antipolar phase is insulating. Similar measurements performed on a nominally identical superlattice grown on a GSO substrate is shown in Extended Data S17.

DFT calculations indicate that the band gaps of the polarization wave and antipolar phases in their hypothetical bulk forms are very close to that of the bulk ground-state *R*3*c* phase (2.15 eV, see Extended Data Fig. 14 and Methods). However, a supercell calculation with a slab of two layers of the polarization wave *Pc* phase alternating with two layers of *Pnma*-AFE $BiFeO_3$ phase frozen in their bulk structures, yields a significant 0.45 eV reduction in the band gap of the polarization wave phase to 1.7 eV (Extended Data Fig. 14c-d). The calculated layer-by-layer and wave-polarization-resolved densities of states (Figs. 3e and f and Extended Data Fig. 14e-g) reveal that the effective band-gap reduction is the result of an alternating band bending caused by the built-in electric fields induced by the local $\pm[001]_{pc}$-oriented components of the wave polarization perpendicular to the net polarization of the phase. This large electrostatic reduction in the effective bandgap is responsible for the significant enhancement in conductivity seen in the MIM results (Supplementary Text 8). Moreover, the large bandwidth of conduction observed in MIM at MHz and GHz frequencies suggests that the conduction is mediated by electrons, and not extrinsic effects, such as oxygen vacancy migration[53]. X-ray absorption spectroscopy obtained via spatially-resolved photoemission electron microscopy on a mixed-phase superlattice also confirms the $Fe^{3+}$ valence state (Extended Data Fig. S24). We note that contrary to previous work on one-dimensional conduction observed at ferroelectric domain walls (DWs)[54,55] and phase boundaries[56], the electron conduction observed here occurs through the bulk of the polar BFO phase and does not rely on the presence of defects (DWs and phase boundaries). Moreover, it is isotropic in nature and is a direct manifestation of the confinement of the polarization wave in a superlattice structure.

**Electric Field Control of Centrosymmetry**

The phase coexistence and its impact on the physical properties lays the foundation for the study of pathways to interconversion between the two phases with an applied electric field. For this purpose, test structures were fabricated by lithographically patterning a set of orthogonal in-plane electrodes that enable the application of a lateral electric field along both $[100]_{pc}$ and $[010]_{pc}$ (Methods and Extended Data Fig. 15). Using these structures, we measured the field-dependent SHG concurrently with in-plane DC transport. Starting in the virgin, mixed-phase state (Fig. 4a), upon applying a lateral electric field along the $[010]_{pc}$, we observe the conversion of the polar phase (strong SHG signal) to the antipolar phase (no SHG signal) in a nonvolatile fashion (Fig. 4b). We emphasize here that, remarkably, an application of an electric (*i.e.* polar) field along the $[010]_{pc}$ eliminates the net polarization and stabilizes the antipolar state. A similar effect is observed with the application of a negative bias along this direction (*i.e.*, along the $[0\bar{1}0]_{pc}$; Extended Data Fig. 11 and Supplementary Text 3). Grey regions within the SHG images denote the lithographically-defined electrodes, where the electric field direction and magnitude are denoted above each SHG panel and overlaid on the electrodes. It is noteworthy that reversing the electric field direction does not revert the antipolar phase back into the polar phase, thus suggesting that either the polar phase has been irreversibly converted into the antipolar phase or that conversion of the antipolar phase back to the polar phase requires an electric field along a different direction. Indeed, by applying an electric field along the $[100]_{pc}$ (Fig. 4c), the antipolar phase fully reverts back into the polar phase with near uniformity and is nonvolatile (> $10^6$ sec, Extended Data Fig. S18). Finally, reapplication of a lateral electric field along the $[010]_{pc}$ reforms a near-uniform antipolar state between the electrodes by driving the polar/antipolar phase boundary back in a nonvolatile fashion (Fig. 4d), highlighting the complete reversibility of the process. We note that while coexistence of antipolar and polar phases and electric field conversion from antipolar to polar states has been reported[57–59], these field-induced phase transitions are typically a volatile saturation of antipolar dipoles along the electric field direction[60,61], relax after relatively short transients, and are not a fundamental change of the ground state of the system. Here, the nonvolatile interconversion between the two phases occurs via a nucleation and growth process, which is illustrated in a sequence of SHG images with finer scale electric field steps (Extended Data Fig. S16, S18, S19, and Supplemental Text 7), that depicts the motion of the phase boundary and the non-volatility of the symmetry transformation and confirms the first-order nature of the phase transformation. The nonvolatile transformation between the phases is also confirmed with correlative SHG, PFM, and MIM imaging (Extended Data Figs. 10, 11, and 17 and Supplemental Text 6, 7, and 8) on separate devices, which also demonstrates repeatability.

These electric field-dependent switching processes are sensitive to the orientation of the electric field with respect to the antipolar phase versus the polar phase. In other words, an electric field applied parallel to the polarization wave direction (i.e., $[100]_{pc}$) results in stabilization of the non-centrosymmetric polar phase, while an electric field applied perpendicular to this favors the centrosymmetric antipolar phase, rather than a reorientation of the polarization in the wave. Fig. 1a plots the DFT-computed energies for these three scenarios (in red and green for the polarization waves and orange for the antipolar phase). While the relative energy difference between the two polar phases – those with perpendicularly-oriented polarization waves – is small and nearly the same as the energy of the antipolar phase, the energy barrier between the polar phases in the two orientations is twice as large as that between the polar phase and the antipolar phase (Fig. 4e), due to the similar octahedral tilt patterns of these two phases (top, Fig 4e). As a result, the centrosymmetric antipolar phase is kinetically favored over the non-centrosymmetric polar phase under a perpendicularly-applied electric field.

Based on the field-dependent imaging of the SHG signal, we record the hysteretic behavior of the SHG intensity as a function of the applied electric field at any given point in the test structure (Fig. 4f; from the location circled in Fig. 4a), where the positive horizontal axis indicates electric field along the $[010]_{pc}$, and the negative horizontal axis denotes electric field along the $[100]_{pc}$. The SHG intensity is manipulated by as much as three orders of magnitude with full non-volatility (stable $>10^6$ sec., Extended Data Fig. S18). We note that, while prior efforts have been dedicated to manipulating the inversion symmetry of materials probed with electric field- and current-driven SHG[32,62–66], as well as geometric patterning[29,67], these effects were typically weaker than those observed here and generally arise from manipulation of electronic band structures, in contrast to overtly manipulating the crystal symmetry through a phase transition, as is shown here. The field-dependent hysteresis in the SHG data is corroborated by the corresponding hysteretic DC transport (Fig. 4g, also see Extended Data Fig. S22), where the DC resistivity between the two metal electrodes is modulated by over four orders of magnitude; consistent with the range of conductivity observed in each phase in the MIM measurements (Fig. 3), and also consistent with an effective bandgap change of ~0.3 eV. The difference in the coercivity between the SHG (Fig. 4f) and resistivity hysteresis (Fig. 4g) is directly attributed to the fact that the SHG is measured at a single point, whereas the resistivity is a macroscopic manifestation of the connectivity between the metal electrodes.

**Outlook**

In summary, we have demonstrated a platform by which we can create mixed polar/antipolar phase coexistence in the BFO system by imposing electrostatic and structural boundary conditions upon it. Furthermore, our results show a reversible pathway to convert from one phase to the other with an external electric field, including, remarkably, a mechanism by which an antipolar state can be stabilized with an electric field. Such symmetry and polar order changes are accompanied by dramatic changes in SHG intensity and, more surprisingly, large changes in both the DC and microwave conductivity. It is noteworthy that the significant changes in the SHG and transport behavior, which are a consequence of such a symmetry phase transition, can enable its use as a mechanism in nonvolatile information storage[56], neuromorphic computing[68], and cross-functional devices which take advantage of optical, electronic, and ferroic properties of materials.

**Methods**

*Molecular Beam Epitaxy:* Superlattices of alternating $TbScO_3$ and $BiFeO_3$ layers are synthesized by reactive oxide molecular-beam epitaxy in a Veeco GEN10 MBE with *in-situ* reflection high-energy electron diffraction (RHEED) and x-ray diffraction (XRD) using distilled ozone as the oxidant species. $(BiFeO_3)_m/(TbScO_3)_n$, where $n$ and $m$ refer to the thickness, in unit cells, of the $BiFeO_3$ and $TbScO_3$, respectively, are grown on $(110)_O$ $TbScO_3$ and $GdScO_3$ substrates, where the subscript o denotes orthorhombic indices; note that for $TbScO_3$ and $GdScO_3$ $(110)_o=(001)_{pc}$. The superlattices are grown at a substrate temperature between 650 °C and 680 °C in a background pressure of $5\times10^{-6}$ Torr (mmHg) of distilled $O_3$ (estimated to be 80% pure $O_3$). Substrate temperatures are measured by an optical pyrometer with a measurement wavelength of 980 nm focused on a platinum layer deposited on the backside of the substrate.

*Electrode Patterning and Sputter Deposition:* The device electrodes are a bilayer of Ta(4 nm)/Pt(40 nm), where the Ta metal is used as an adhesion layer. The metals were deposited using DC magnetron sputtering (AJA International) at a nominal room temperature with an Ar sputter gas pressure of 2 mTorr and a background base pressure of $\sim3\times10^{-8}$ Torr. Deposition rates for each element were calibrated using X-ray reflectivity measurements of the film thickness. All electrode devices were patterned using a mask-less aligner (Heidelberg Instruments) and standard lift-off processes.

*Scanning Transmission Electron Microscopy (STEM):* The plan-view and cross-sectional STEM samples of the $[(BiFeO_3)_{14}/TbScO_3)_{10}]$ superlattices were prepared using a FEI Helios focused ion beam (FIB) with a final milling step of 2 keV to reduce damage. The initial sample surface was protected from ion-beam damage by depositing carbon and platinum layers prior to milling. HAADF-STEM images were recorded using a Cs-corrected ThermoFisher Scientific Spectra300 operated at 300 kV, with beam semi-convergence angle of 30 mrad and beam current of 15 pA.

*Scanning convergent beam electron diffraction (SCBED) for polarization mapping:* We performed SCBED experiments using a next-generation electron microscopy pixel array detector (NG-EMPAD). SCBED works by rastering a focused probe in two-dimensional (2D) real space ($x$,$y$) and collecting a full 2D CBED pattern ($k_x$, $k_y$) at each probe position, resulting in a 4D dataset[69,70]. Experimental data was acquired using a FEI Titan operated at 300 keV with 10 pA beam current, 2.45 mrad semi-convergence angle, having a probe of ~8 Å FWHM (full-width at half-maximum). The CBED patterns were captured by the EMPAD with exposure time set to 100 μs per frame, for which a 512 × 512 scan can be recorded in under 2 minutes. Due to dynamical diffraction effects, the charge redistribution due to ferroelectric polarization leads to intensity asymmetry in Friedel pairs of Bragg reflections. We employ Kikuchi bands as a more robust means to extract polarity information against crystal mis-tilts, which often occurs in ferroic oxides due to disinclination strain. By matching with dynamical diffraction simulations, we can unambiguously determine the polarization directions in real space[42–44].

*Laboratory-based X-ray diffraction:* Structural characterization was performed using a Panalytical X'Pert3 MRD 4-circle diffractometer with a Cu source. Two-dimensional reciprocal space maps were measured around the $TbScO_3$ (220) peak.

*Ferroelectric Characterization:* Study of the superlattice domain structures was carried out using an atomic force microscope (MFP-3D, Asylum Research). Dual AC resonance tracking piezoforce microscopy (PFM) was conducted using a conductive Pt-coated probe tip (MikroMasch HQ:NSC18/PT) to image the ferroelectric domain structures using lateral and out-of-plane imaging modes. In-plane hysteresis loops were taken using a Radiant Technologies Precision Multiferroic Tester with a frequency of 100 Hz, and an applied voltage of 150 V on patterned interdigitated electrodes (See Fig. S15). The lateral resistivity (ρ) is estimated from the two-point resistance (R) measurement shown in Extended Data Fig. S22 via $\rho=R*L/A$, where L=6 μm is the distance between two electrodes and A is the cross-sectional area of the sample where the electric field is applied. We approximate that the electric field is uniform throughout the thickness of the sample. The resistance is always measured across the electrodes along the $[100]_{pc}$ direction, as shown in the images below and now added to the Extended Data. However, we note that the results are similar if the resistance is measured across the electrodes along the $[010]_{pc}$. In other words, the resistivity is isotropic.

*Microwave Impedance Microscopy (MIM):* The MIM experiments were carried out on a commercial AFM platform (XE-70, Park AFM). The electrically shielded microwave cantilever probes are commercially available from PrimeNano, Inc[49]. The two output channels of MIM correspond to the real and imaginary parts of the local sample admittance, from which the effective ac conductivity of the sample can be deduced. Numerical simulation of the MIM signals was performed by the FEA software COMSOL4.4. Details of the FEA are included in Fig. S12 and Supplementary Text 4.

*Second Harmonic Generation (SHG):* SHG measurements were carried out in a normal-incidence, reflection-geometry. A Ti/Sapphire oscillator was used for light excitation with ~100 fs pulses and center wavelength of 900 nm, a 78 MHz repetition rate, and an average power of <1 mW. To arbitrarily control the polarization of the incoming light, we use a Glan-Thompson polarizer and subsequently send the light through a half-waveplate. The polarized light was then sent through a short pass dichroic mirror and focused on the sample using an oil immersion objective lens (OL, NA = 1.4). The back-scattered SHG signal was sent through a short pass filter and detected using a spectrometer (SpectraPro 500i, Princeton Instruments) with a charge-coupled device camera (Peltier-cooled CCD, ProEM+: 1600 eXcelon3, Princeton Instruments). A linear polarizer on the back-end optics was used to select emitted light polarization for detection. Diffraction-limited confocal scanning microscopy was used to create SHG intensity maps. A commercial Thorlabs polarimeter was used at the sample location to confirm the incoming light polarization incident on the sample, as well as the light polarization entering the detector. All SHG maps shown throughout the manuscript were performed using $[100]_{pc}$-polarized incident light (light polarized along the polarization wave orientation) with no polarizer on the back-end optics. *In-situ*

DC electric fields are applied to the sample using a Keithley 6430 SUB-FEMTOAMP SourceMeter via a custom, shielded printed circuit board.

*SHG Selection Rules:* We are working with light incident along the direction $z \sim [001]_{pc}$ so that only the $x$ and $y$ components of the SHG susceptibility tensor are addressed. For the polar point group m with $x \sim [100]_{pc}$ (i.e. spontaneous polarization along $x$) this leads to the susceptibilities $\chi_{yyy}$, $\chi_{yxx}$, $\chi_{xxy} = \chi_{xyx}$, $\chi_{xyy}$, $\chi_{yyxy} = \chi_{yxy}$ that are expected to be nonzero and can be probed by our experiment. These susceptibilities were used to fit the SHG polarity data in Fig. 2c and 2d.

*First-Principles Calculations:* Calculations were performed using DFT[71] with the projector augmented wave (PAW) method[72] as implemented in the Vienna *ab initio* simulation package (VASP 5.4.4)[73]. We used a 12x12x12 k-point Γ-centered mesh to sample the Brillouin zone corresponding to a five-atom unit cell and chose an energy cutoff of 850 eV for the plane-wave basis. The following valence electron configurations were used: $6s^2 6p^3$ for bismuth, $3d^7 4s^1$ for iron, and $2s^2 2p^4$ for oxygen. Note that the inclusion of the $5d^{10}$ electrons in the valence manifold for bismuth and the $3p^6$ for iron were tested for bulk structures and gave similar results. The PBEsol + U functional form of the generalized gradient approximation[74] was used, with a commonly used value of $U_{eff} = 4$ eV for the Fe 3d orbitals[75,76], according to Dudarev's approach[77]. No ionic relaxation was carried out in the supercell in which the DOS were computed. The energy barrier calculations were performed by interpolating the ionic positions and volumes linearly between the starting and ending structures and relaxing the electronic structure at fixed ionic positions.

*Phase Field Simulations:* In the phase-field method, the local free energy density is expressed as a function of the local polarization $P_i$ ($i$ =1-3), local oxygen octahedral tilt order (OTs) $\theta_i$ ($i$ =1-3) and antiferroelectric order parameter (AFEs) q$i$ ($i$ =1-3). The total free energy of a mesoscale domain structure described by the spatial distribution of polarization, oxygen octahedral tilt and antiferroelectric order is then the volume integration of bulk free energy density, elastic energy density, electrostatic energy density and gradient energy density,

$$F = \int \Big[\alpha_{ij}P_iP_j + \alpha_{ijkl}P_iP_jP_kP_l + \alpha_{ijklmn}P_iP_jP_kP_lP_mP_n + \beta_{ij}\theta_i\theta_j + \beta_{ijkl}\theta_i\theta_j\theta_k\theta_l + \gamma_{ij}q_iq_j + \gamma_{ijkl}q_iq_jq_kq_l + \gamma_{ijklmn}q_iq_jq_kq_lq_mq_n + t_{ijkl}P_iP_jq_kq_l + f_{ijkl}P_iP_j\theta_k\theta_l + h_{ijkl}q_iq_j\theta_k\theta_l + \frac{1}{2}g_{ijkl}P_{i,j}P_{k,l} + \frac{1}{2}k_{ijkl}\theta_{i,j}\theta_{k,l} + \frac{1}{2}m_{ijkl}q_{i,j}q_{k,l} + \frac{1}{2}c_{ijkl}(\varepsilon_{ij} - \varepsilon_{ij}^0)(\varepsilon_{kl} - \varepsilon_{kl}^0) - E_iP_i - \frac{1}{2}\varepsilon_b\varepsilon_0E_iE_j\Big]\, dV,$$

where $\alpha_{ij}$, $\alpha_{ijkl}$, $\alpha_{ijklmn}$, $\beta_{ij}$, $\beta_{ijkl}$, $\gamma_{ij}$, $\gamma_{ijkl}$, $\gamma_{ijklmn}$, $t_{ijkl}$, $f_{ijkl}$ and $h_{ijkl}$ are local potential coefficients representing the stiffness with respect to the changes in polarization, oxygen octahedral tilt and antiferroelectric order. $g_{ijkl}$, $k_{ijkl}$ and $m_{ijkl}$ are the gradient energy coefficients of polarization, OTs and AFEs, respectively. $\varepsilon_b$ is the isotropic background dielectric constant and $\varepsilon_0$ is the dielectric constant of free space. The eigenstrain $\varepsilon^0$ + is coupled to polarization, AFEs and OTs through $\varepsilon_{ij}^0 = Q_{ijkl}P_iP_l + N_{ijkl}q_iq_l + L_{ijlk}\theta_i\theta_j$, where $Q_{ijkl}$, $N_{ijkl}$ and $L_{ijkl}$ are the coupling coefficients.

The temporal and spatial evolution of polarization, OTs and AFEs are governed by the relaxation equations leading to the minimization of the total free energy of the system. In the simulations, periodic boundary conditions are employed along three dimensions. For the mechanical boundary condition, the in-plane directions are clamped while the out-of-plane direction is assumed to be stress-free. A pseudo-2-D mesh of $300 \times 2 \times N$ is used, where N indicates the film thickness and the grid spacing is 0.4 nm. The value of N ranges from 60 to 345 based on different simulation conditions. Paraelectric insulating layers are simulated with different dielectric constants. All simulations are performed for a temperature of 300 K.

**Acknowledgements**

R.R., L.W.M., D.A.M., and D.G.S. acknowledge support from the Army Research Office under the ETHOS MURI via cooperative agreement W911NF-21-2-0162. The MIM work (J.Y., D.L., and K.L.) was supported by the Office of Science, Office of Basic Energy Sciences, of the U.S. Department of Energy under Contract No. DE-SC0019025. Computational resources were provided by ETH Zürich and the Swiss National Supercomputing Center (CSCS), Project ID No. s889. Work at ETH was supported by ETH Zürich and the Körber Foundation. Work at the Molecular Foundry was supported by the Office of Science, Office of Basic Energy Sciences, of the U.S. Department of Energy under Contract No. DE-AC02-05CH11231. L.C. acknowledges financial support from the Ford Foundation and the University of California President's Postdoctoral Fellowship Program. YT.S. and D.A.M acknowledge financial support from the Department of Defense, Air Force Office of Scientific Research under award FA9550-18-1-0480. The electron microscopy studies were performed at the Cornell Center for Materials Research, a National Science Foundation (NSF) Materials Research Science and Engineering Centers program (DMR-1719875). The microscopy work at Cornell was supported by the NSF PARADIM NSF-MRI-1429155, with additional support from Cornell University, the Weill Institute and the Kavli Institute at Cornell. The authors acknowledge fruitful discussions regarding diffraction imaging with Prof. Jian-Min Zuo as well as M.

Thomas, J. G. Grazul, M. Silvestry Ramos, K. Spoth for technical support and careful maintenance of the instruments. The authors thank Xiaoxi Huang and Abel Fernandez for fruitful conversations and Prof. Megan E. Holtz for preliminary electron microscopy studies.

**Author Contributions**

R.R., D.G.S., and L.C. conceived the project and planned the experiments; YT.S., H.K.P., and M.E.H. performed transmission electron microscopy (TEM), TEM sample preparation, and atomically-resolved polar and structural analysis under supervision of D.A.M.; A.B.M. optimized synthesis of the superlattices under supervision of D.G.S; L.C. and P.B. performed in-situ SHG measurements with help from A.R. and E.B, and M.F.; L.C., P.B, and E.B. prepared the experimental SHG setup; P.B. performed PFM imaging under supervision from L.C.; L.C. and P.B. performed electronic transport measurements. J.Y. and D.L. performed microwave impedance microscopy and analysis with supervision from K.L. M.M. performed lab-based x-ray structural characterization and analysis; L.C. and E.P designed and microfabricated the electric field devices. L.C. deposited metal layers; First-principles calculations were performed by B.F.G. under supervision of N.S.; Phase field calculations were performed by C.D. and F.X. under the supervision of L.-Q. C.; SHG analysis was completed by L.C., P.B, and M.F.; Scandate crystal substrates were grown by S.G.; L.C., Y.T., R.R, K.L., and L.W.M. wrote the manuscript.

**Conflict of Interest**

K.L. holds a patent on the MIM technology, which is licensed to PrimeNano, Inc., for commercial instruments. The terms of this arrangement have been reviewed and approved by the University of Texas at Austin in accordance with its policy on objectivity in research. The remaining authors declare no conflict of interest.

**Additional Information**

Correspondence and request for materials should be addressed to R.R. and L.C.

**Data Availability**

The data supporting the findings of this study are available within the paper and other findings of this study are available from the corresponding authors upon reasonable request.

# Figures

## Figure 1

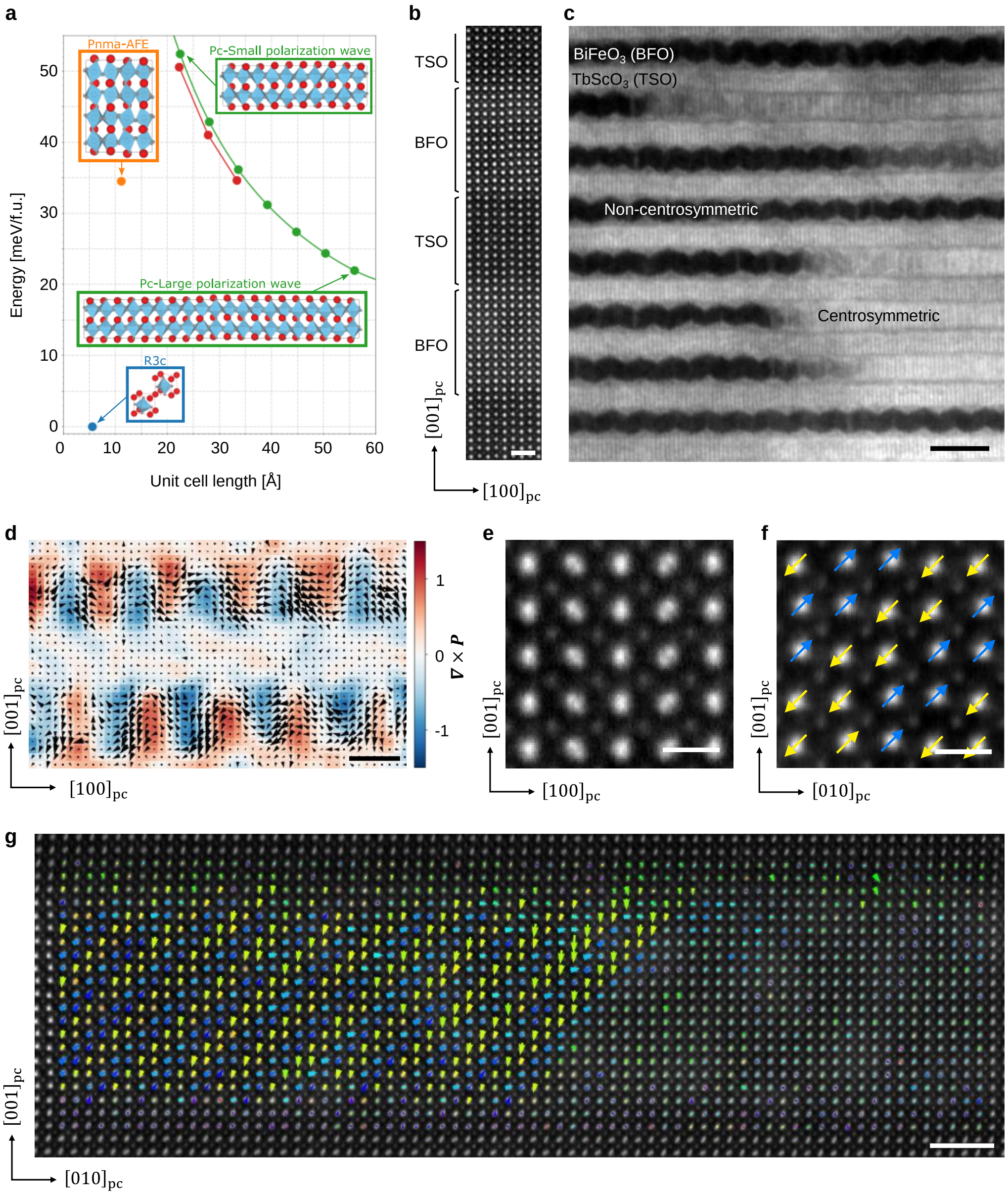

**Figure 1 | Mixed-phase coexistence of polar and antipolar phases in the BFO/TSO superlattice. a)** Structures, energies, and lattice constants of low-energy phases of BFO with zero out-of-plane net polarization predicted using DFT. **b)** Atomically-resolved HAADF-STEM image of the $[(BFO)_{14}/(TSO)_{10}]_{20}$ superlattice along the $[010]_{pc}$ zone axis showing atomically sharp interfaces. Scale bar is 10 nm. **c)** Bright-field image obtained from the SCBED dataset showing the spatial distribution of polar (*Pc*) and anti-polar (*Pnma*-AFE) BFO phases. Scale bar is 10 nm. **d)** Polarization map of the polar phase overlaid with its curl ($\nabla \times \vec{P}$) obtained by analyzing the Kikuchi bands recorded in the SCBED dataset using an EMPAD. Scale bar is 5 nm. Atomic resolution HAADF-STEM images of the **e)** anti-polar phase along $[010]_{pc}$ (left) and **f)** $[100]_{pc}$ (right) zone axes. Scale bar is 5 Å. **g)** HAADF-STEM image along $[100]_{pc}$ showing the atomically sharp interface between the antipolar and polar phases. The vectors represent the displacement of Bi atomic columns relative to the four neighboring Fe columns, showing the "up-up/down-down" distortion along the 45º in the antipolar region. Scale bar is 2 nm.

**Figure 2**

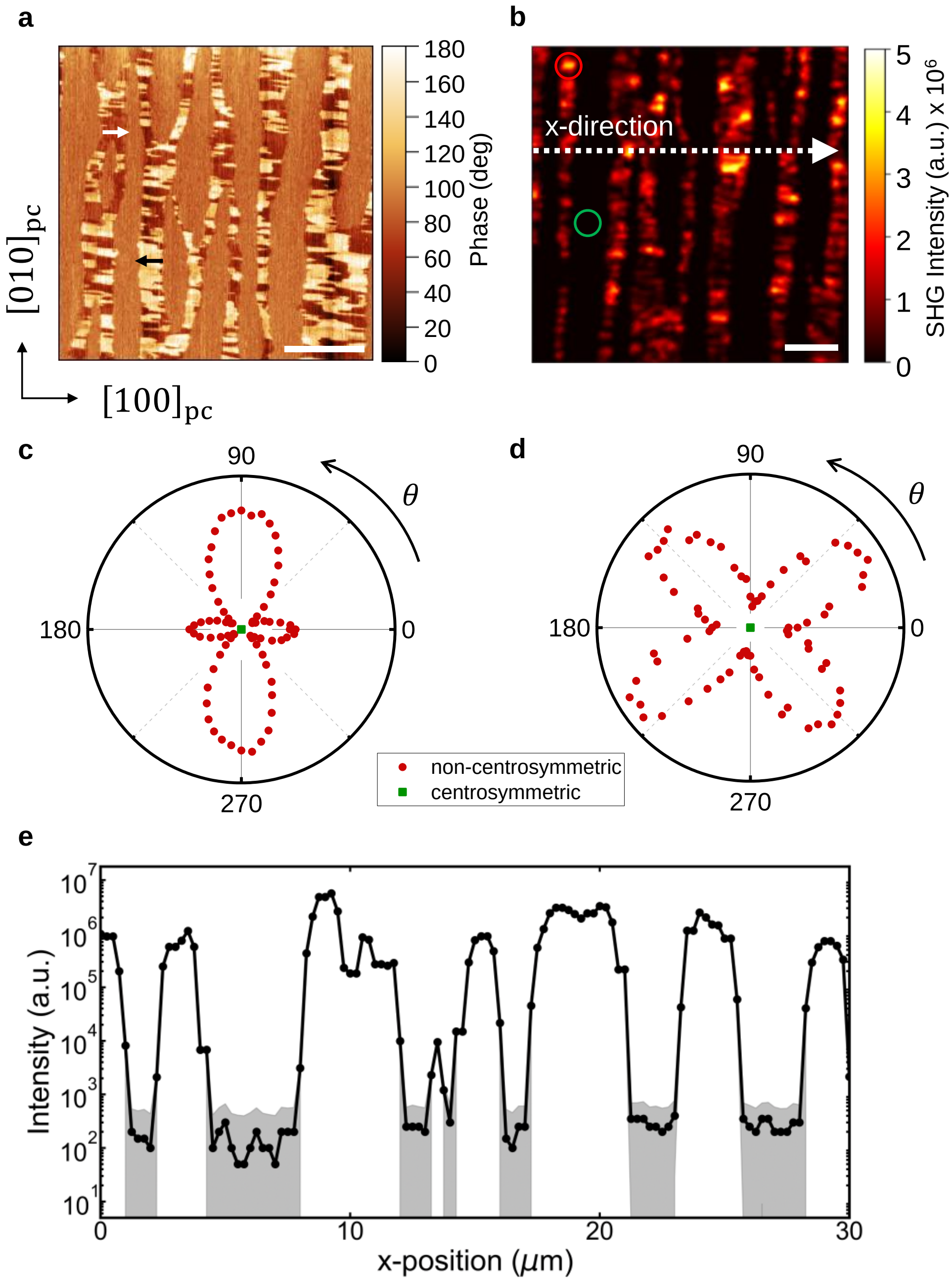

**Figure 2 | Piezoforce and nonlinear optical response of mixed-phase coexistence. a)** PFM phase image of the $[(BFO)_{14}/(TSO)_{10}]_{20}$ superlattice. Dark regions with near 0° phase angle denote net polarization along the $[100]_{pc}$ direction, as indicated by the white arrow, whereas light regions with near 180° phase denote net polarization along the $[\bar{1}00]_{pc}$ direction, as indicated by the black arrow. Areas exhibiting 90° phase have no net polar order and reflect antipolar regions. **b)** Confocal SHG map on a nearby area of the same sample. Areas showing high (low) SHG intensity contain (non-)centrosymmetric BFO. Local SHG polar plots from the red and green circled regions in (b) are shown in (c) and (d), where the incident light polarization (θ) is varied and the corresponding **c)** vertically or **d)** horizontally polarized emitted light at the second harmonic is analyzed. Red and green regions correspond to non-centrosymmetric and centrosymmetric phases, respectively. **e)** SHG

intensity line profile along the dashed white line in (b). The uncertainty (grey bands) come from the noise level of the SHG detector. Scale bars are 5 μm. a.u., arbitrary units.

**Figure 3**

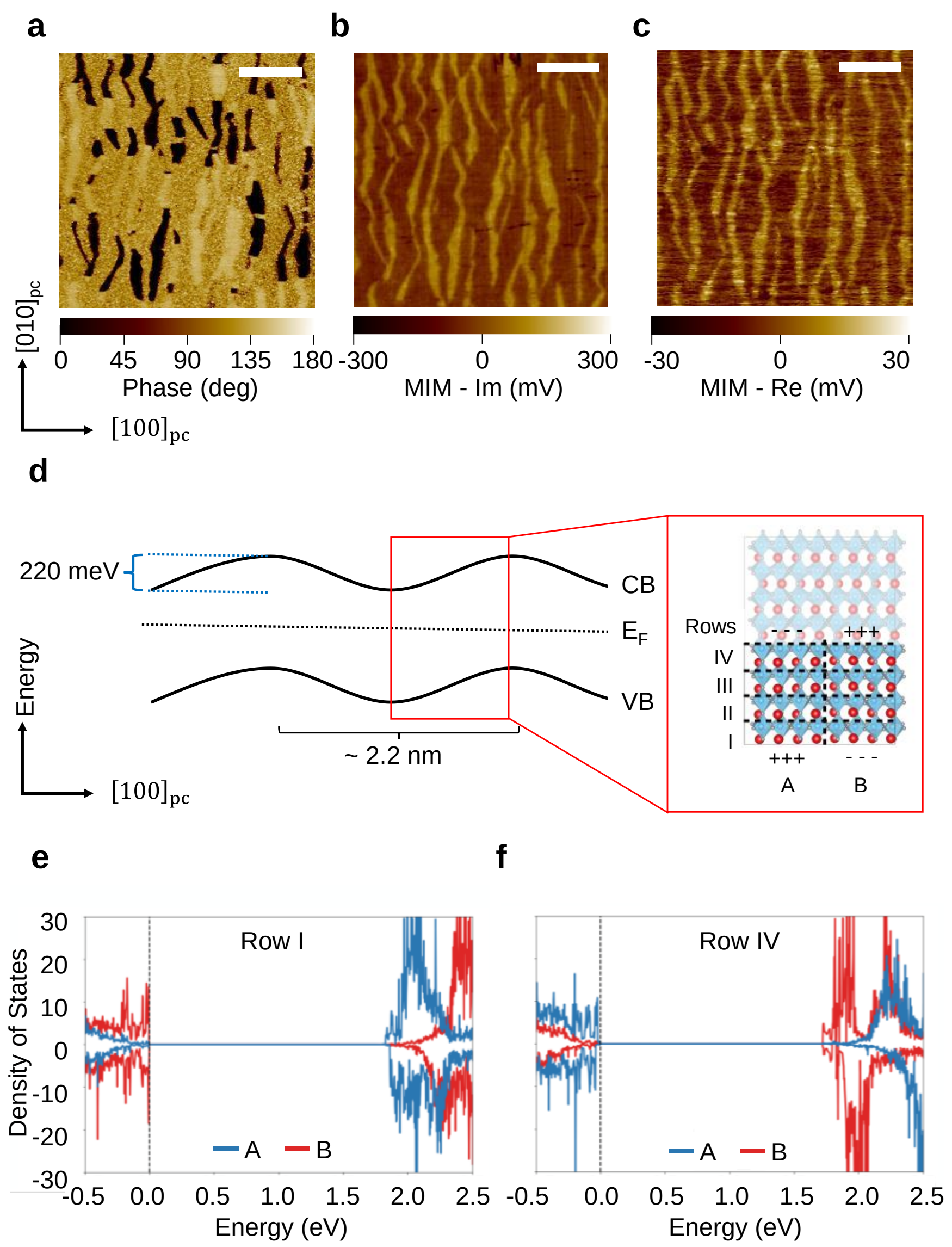

**Figure 3 | Microwave impedance microscopy (MIM) and DFT band structure calculations. a)** PFM phase image of a TSO//[(BFO)$_{14}$/(TSO)$_{10}$]$_{20}$ superlattice, where the contrast is sensitive to polarization along the [100]$_{pc}$ direction. **b)** Imaginary (Im) and **c)** real (Re) parts of the MIM images taken at the same area as (a). Scale bars in (a-c) are 2 μm. **d)** Band bending model of the conduction and valence bands, reducing the effective bandgap of the system by ~0.45 eV. Red "zoomed-in" box shows a labeled real-space schematic of the modeled polarization wave. **e,f)** DFT-computed density of states for a confined polarization wave heterostructure with 2.2 nm periodicity in the two regions indicated in (d), where a built-in voltage is established by [001]$_{pc}$-oriented component of the wave polarization.

**Figure 4**

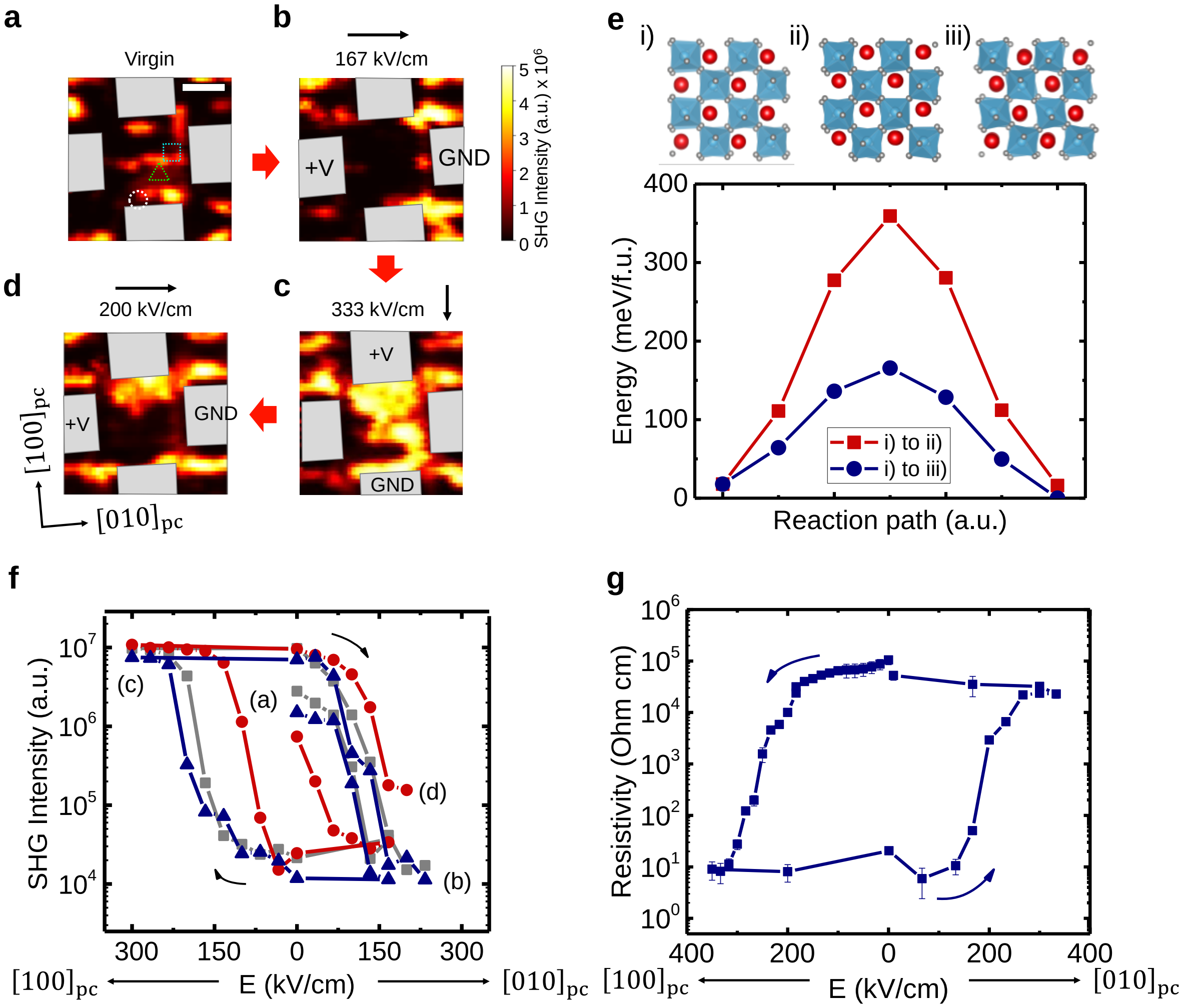

**Figure 4 | Nonvolatile electric field manipulation of SHG intensity and resistivity. a-d)** A series of SHG maps under sequentially applied orthogonal in-plane electric fields. Electrodes are denoted by grey overlaid boxes, where the active region of the sample is located between the four electrodes. The magnitude and orientation of the electric field are denoted by the arrows and text above each panel and on each electrode. Red arrows between each panel denote the sequence of applied electric field. **e) (top)** Structures of i), ii) the polar wave phases with perpendicular polarization orientations, as well as iii) the antipolar phase. Note the differences in the tilt patterns between the two polar phases; the tilt pattern of the antipolar phase is a close match to that of the initial polar phase. (bottom) Energy barriers computed by linearly interpolating between the different phases. **f)** An SHG-electric field hysteresis loop for the regions outlines in (a) showing an approximately three orders of magnitude change in SHG intensity. Each plotted symbol type (circle, square, triangle) corresponds to local hysteresis of the same shape in (a). The positive (negative) horizontal axis denotes electric field oriented along the $[010]_{pc}$ ($[100]_{pc}$) direction. SHG uncertainty corresponds to the background noise of the detector and is smaller than the data points. **g)** Hysteresis of DC resistivity with electric field across the top and bottom electrodes in the device. The positive (negative) horizontal axis denotes electric field oriented along the $[010]_{pc}$ ($[100]_{pc}$) direction. Error bars in the resistivity denote uncertainty of the leakage current during the measurement. Scale bar is 3 μm.

**Supplementary Information:**
**Nonvolatile Electric Field Control of Inversion Symmetry**

**Extended Data**

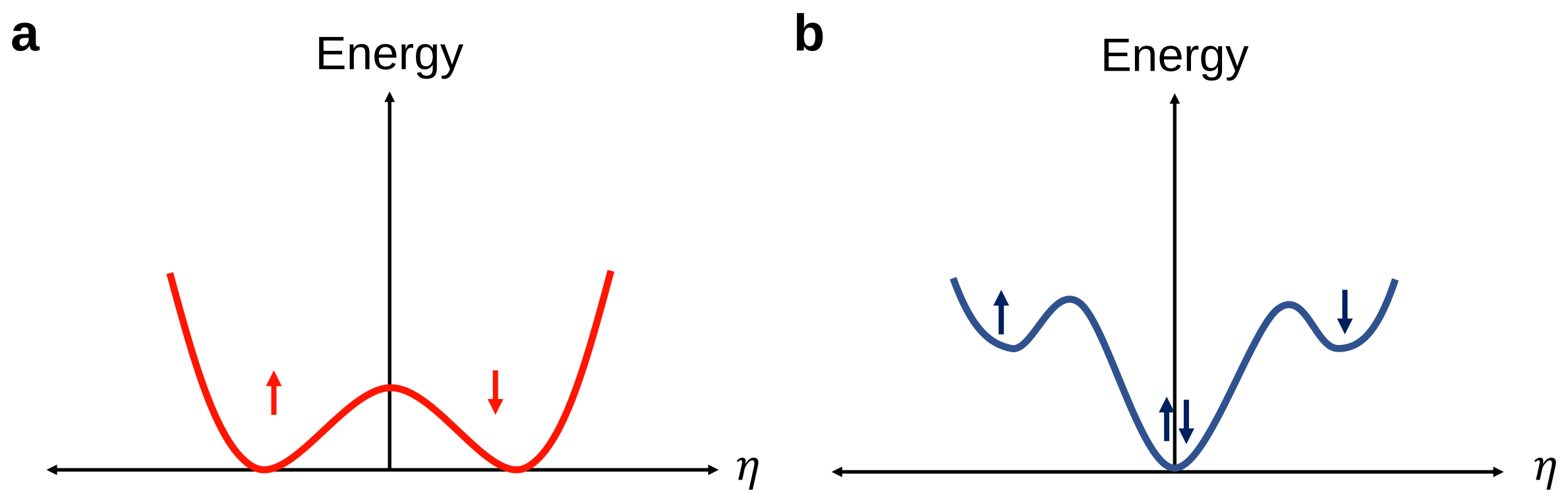

**Figure S1** | Energy-order parameter ($\eta$) landscapes for **a)** ferroelectric and **b)** antiferroelectric systems[1]. Arrows schematically depict polar orientation at each local energy minima.

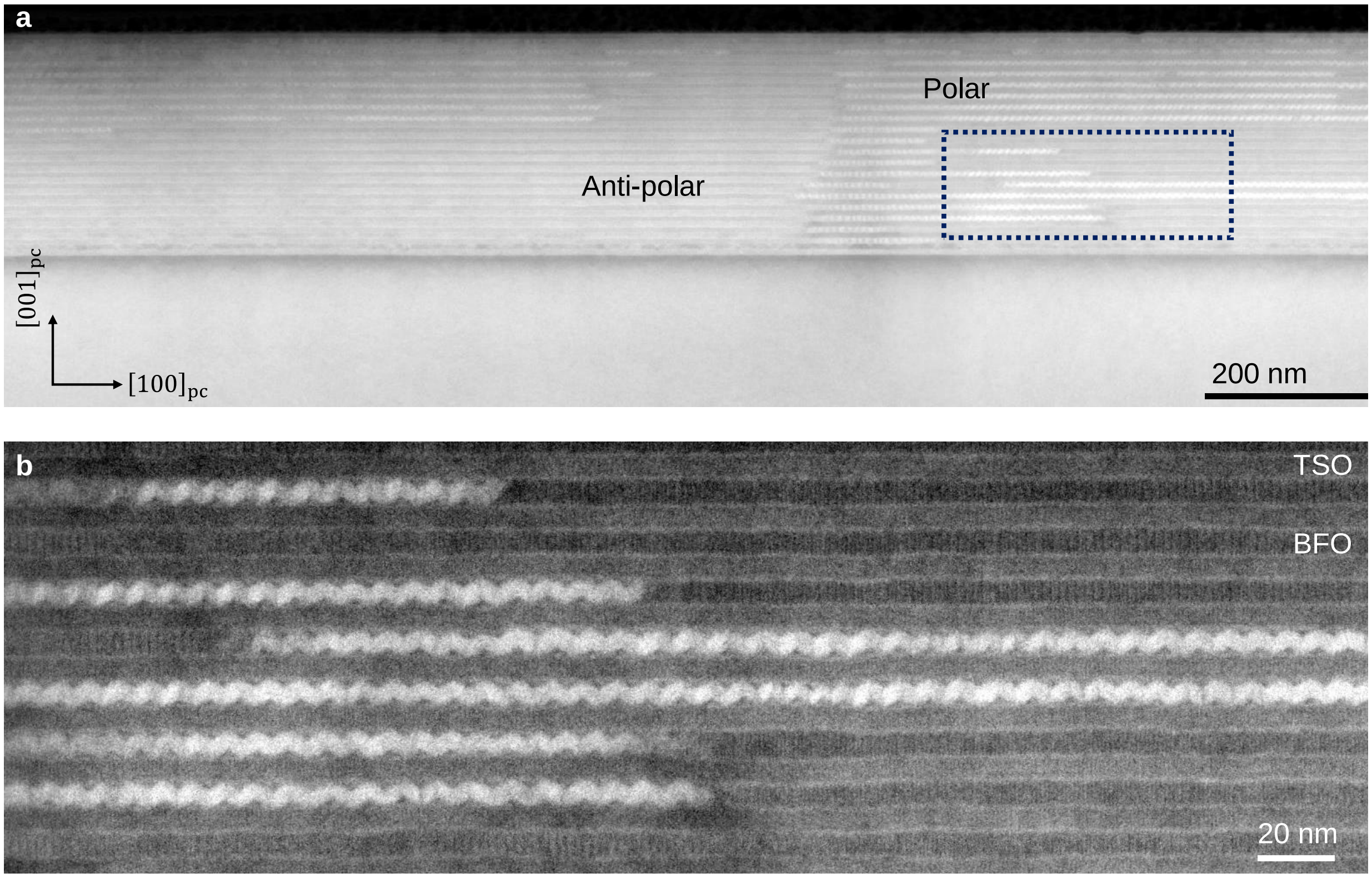

**Figure S2 | Coexisting polar and antipolar phases in the BFO/TSO superlattice. a)** HAADF-STEM image of the $[(BFO)_{14}/(TSO)_{10}]_{20}$//GSO superlattice along the $[010]_{pc}$ zone axis spanning a field of view of ~1.6 µm. To enhance the diffraction contrast, HAADF-STEM was acquired with 300 keV electrons and a probe semi-convergence angle of 2.45 mrad. **b)** Zoomed-in region from the dashed box in (a). The bright regions indicate the polar BFO phase, while the dark regions with vertical stripes are the BFO anti-polar phase.

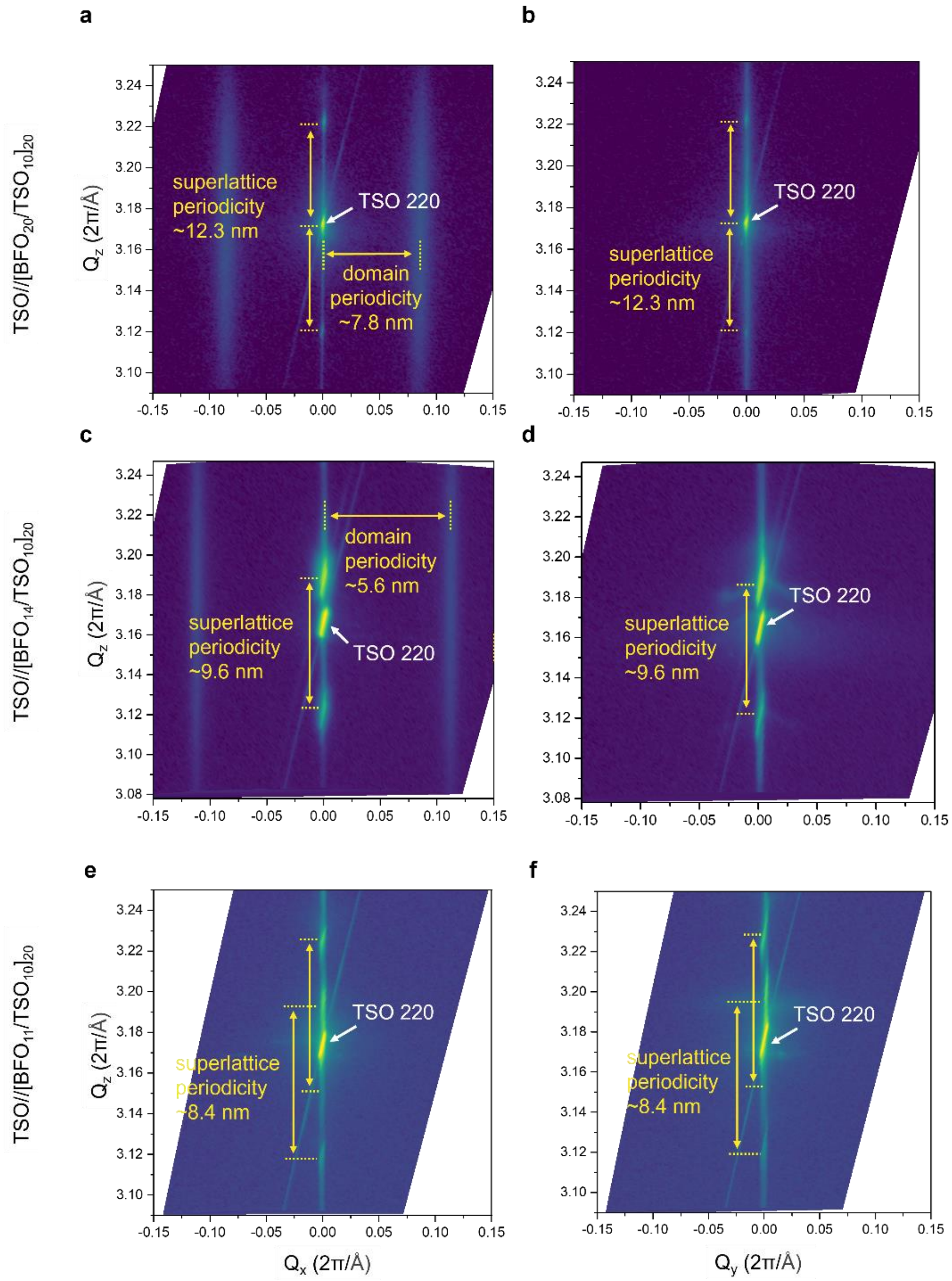

**Figure S3 | Reciprocal space maps (RSMs) of TSO//[(BFO)$_n$/(TSO)$_{10}$]$_{20}$ superlattices.** $Q_z$ ([001]$_{pc}$) versus $Q_x$ [100]$_{pc}$ and $Q_y$ [010]$_{pc}$ reciprocal space maps for various [(BFO)$_n$/(TSO)$_{10}$]$_{20}$ centered around the [220]$_o$ $TbScO_3$ (TSO) substrate peak. **a, b)** when $n$ = 20 unit cells, peaks along the $Q_z$ axis show the out-of-plane periodicity of the superlattice with clearly visible satellite peaks along the **a)** $Q_x$ direction, corresponding to the wave-like periodicity (7.8 nm), which are not present along the **b)** $Q_y$ direction. The diffuse satellite peak is indicative of a lack of correlation along the out-of-plane ($Q_z$) direction. **c,d)** A similar set of RSMs is shown for an $n$ = 14 unit cell superlattice, where a fainter satellite peak (5.6 nm) along the **c)** $Q_x$ is observed, suggesting a smaller fraction of the polarization wave phase, with no clear periodicity in the **d)** $Q_y$ direction. When $n$ = 11 unit cells, no satellite peaks are present along either the **e)** $Q_x$ or **f)** $Q_y$ directions, indicating that no polarization wave phase is present in the film.

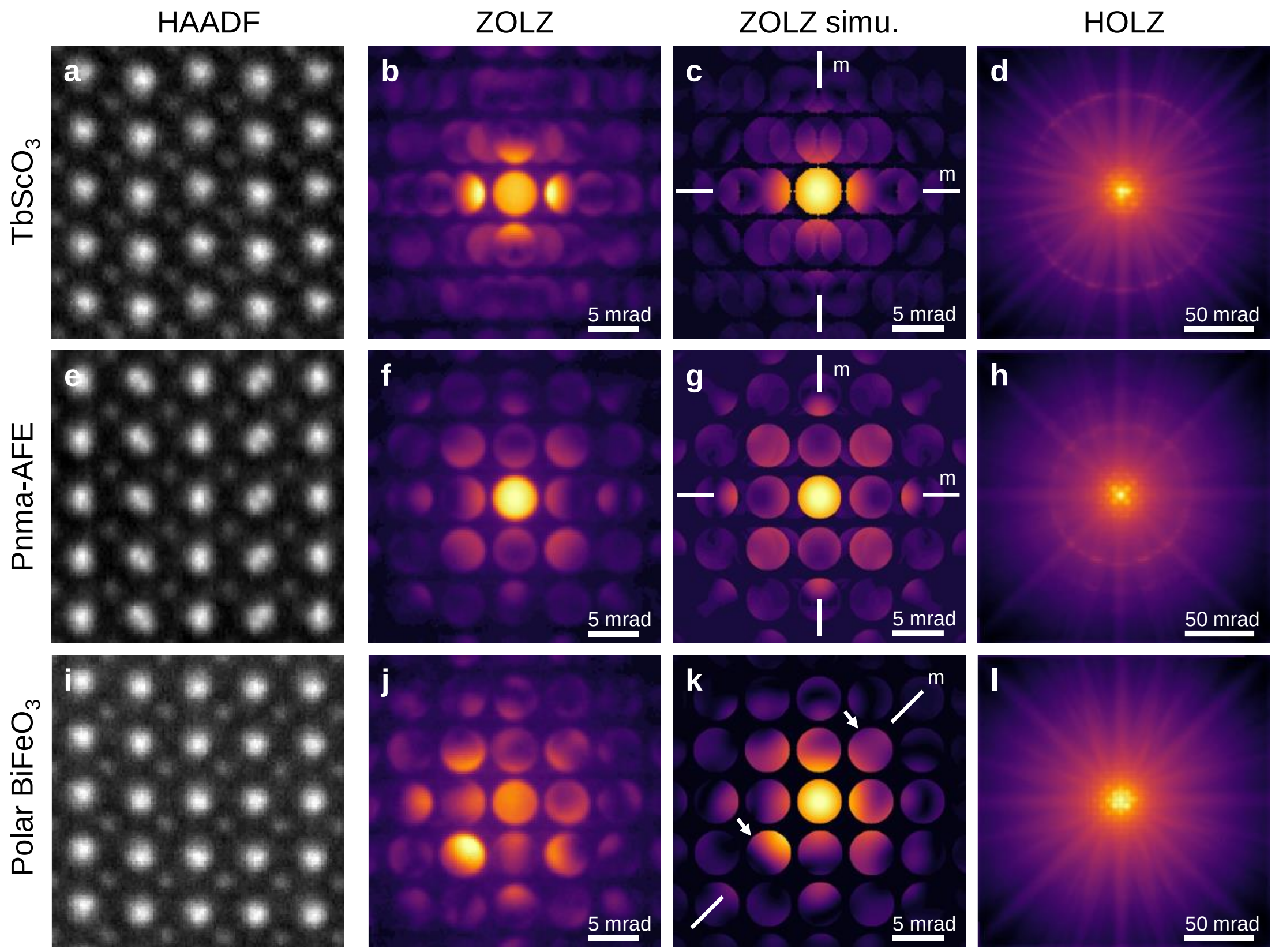

**Figure S4 | Probing local crystallographic symmetry for various phases in the $(BFO_{14}/TSO_{10})_{20}$ superlattice.** Three distinct phases were found, which are the $TbScO_3$ with *Pnma* space group (a-d), $BiFeO_3$ *Pnma*-AFE antipolar phase (e-h) and $BiFeO_3$ polar phase (i-l), respectively. The crystallographic symmetries were determined using a combination of atomic-resolution HAADF-STEM (a,e,i) as well as ZOLZ (zero-order Laue zone, b,f,j) and HOLZ (high-order Laue zone, d,h,l) in CBED patterns recoded using an EMPAD. Both AFE and TSO have the centrosymmetric space group of *Pnma*, which implies that the CBED ZOLZ pattern would exhibit *2mm* symmetry along the [110] zone axis, with mirror planes indicated in simulated CBED patterns (c,g). While ZOLZ is useful for determining projected point group symmetries, the HOLZ rings give 3D information about the unit cell. For example, the HOLZ ring in (d) indicate a unit cell doubling along the projection direction, which is consistent with the known TSO structure. The second HOLZ ring in (h) with $\sim\frac{1}{\sqrt{2}}\times$ radius indicates an additional doubling of the unit cell, which is consistent with the 4x large unit cell of the Pnma-AFE structure. For the polarization wave phase, we observed the breaking of Friedel symmetry which can be attributed to dynamical diffraction effects arising from the ferroelectric polarization. CBED simulations were carried out using the Bloch-wave method, with input structures of TSO (c, *Pnma*), antipolar BFO (g, *Pnma*-AFE) and polar BFO (k, *R3c*), respectively. Note the intensity asymmetry along the mirror plane in (k), as indicated by two white arrows.

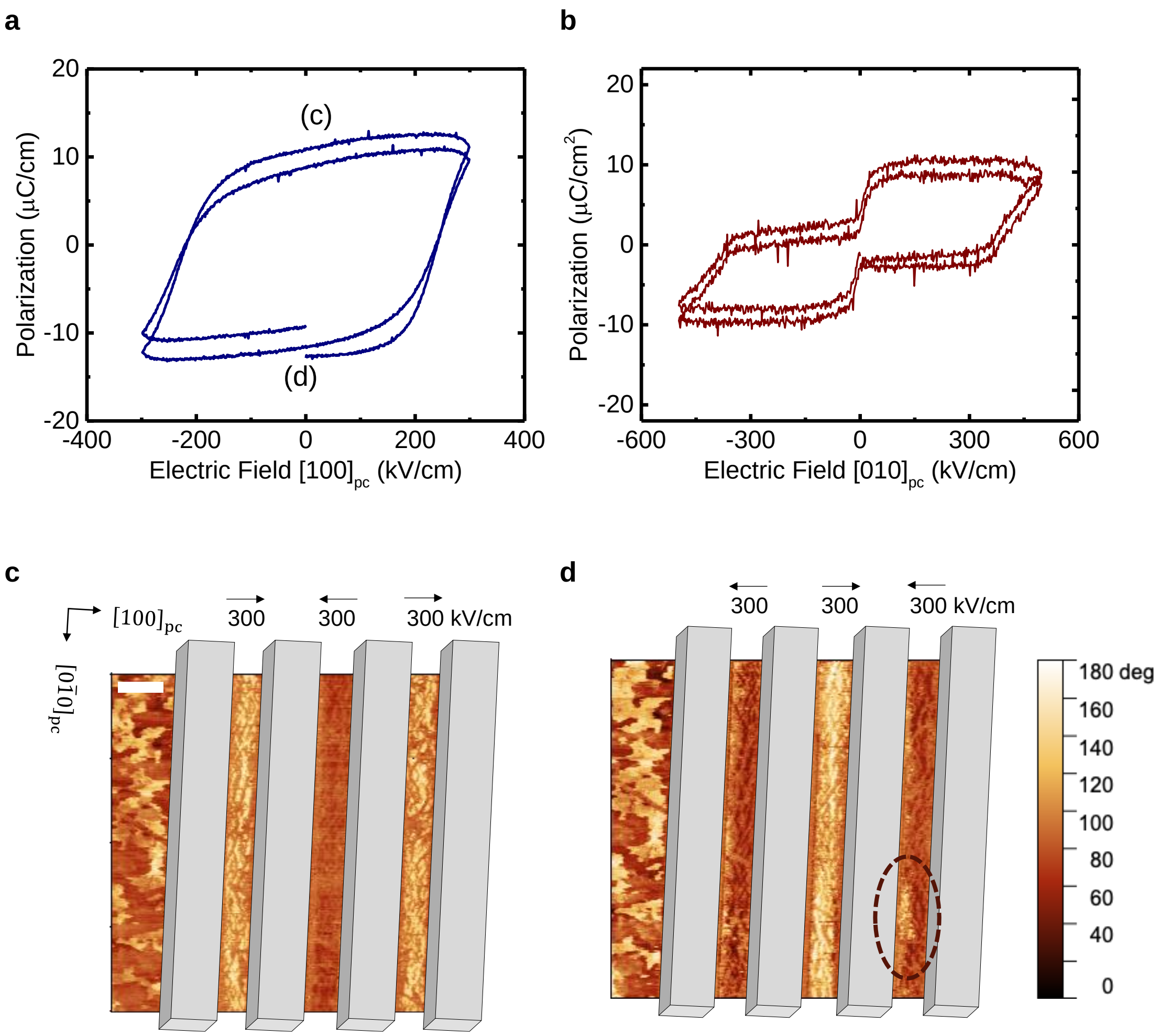

**Figure S5 | Charge-voltage response of the polar and antipolar BFO phases. a)** In-plane hysteresis loops of the TSO//$[(BFO)_{20}/(TSO)_{10}]_{20}$ superlattice, which contain the pure ferroelectric wave phase of BFO. **b)** In-plane hysteresis loops of the TSO//$[(BFO)_{11}/(TSO)_{10}]_{20}$ superlattice, which contains the pure antipolar phase of BFO. Near zero remnant polarization and two-step switching are indicative of antiferroelectric behavior, in which the antipolar state is saturated into a polar state at high electric field. **c)** and **d)** show PFM phase images of the remnant state of the ferroelectric device ($[(BFO)_{20}/(TSO)_{10}]_{20}$) poled in opposite directions, as indicated in (a). The grey, three-dimensional boxes denote the interdigitated electrode (IDE) placement, with the electric field magnitude and direction shown above each region of the device. The region to the left of the IDEs remains unpoled in its virgin state and is used for reference. Regions of oppositely-oriented electric field have an ~180° phase shift in the PFM image, indicating a 180° reversal of the wave polarization. The dashed circle in (d) highlights a partially switched area during poling. Scale bar is 2 μm.

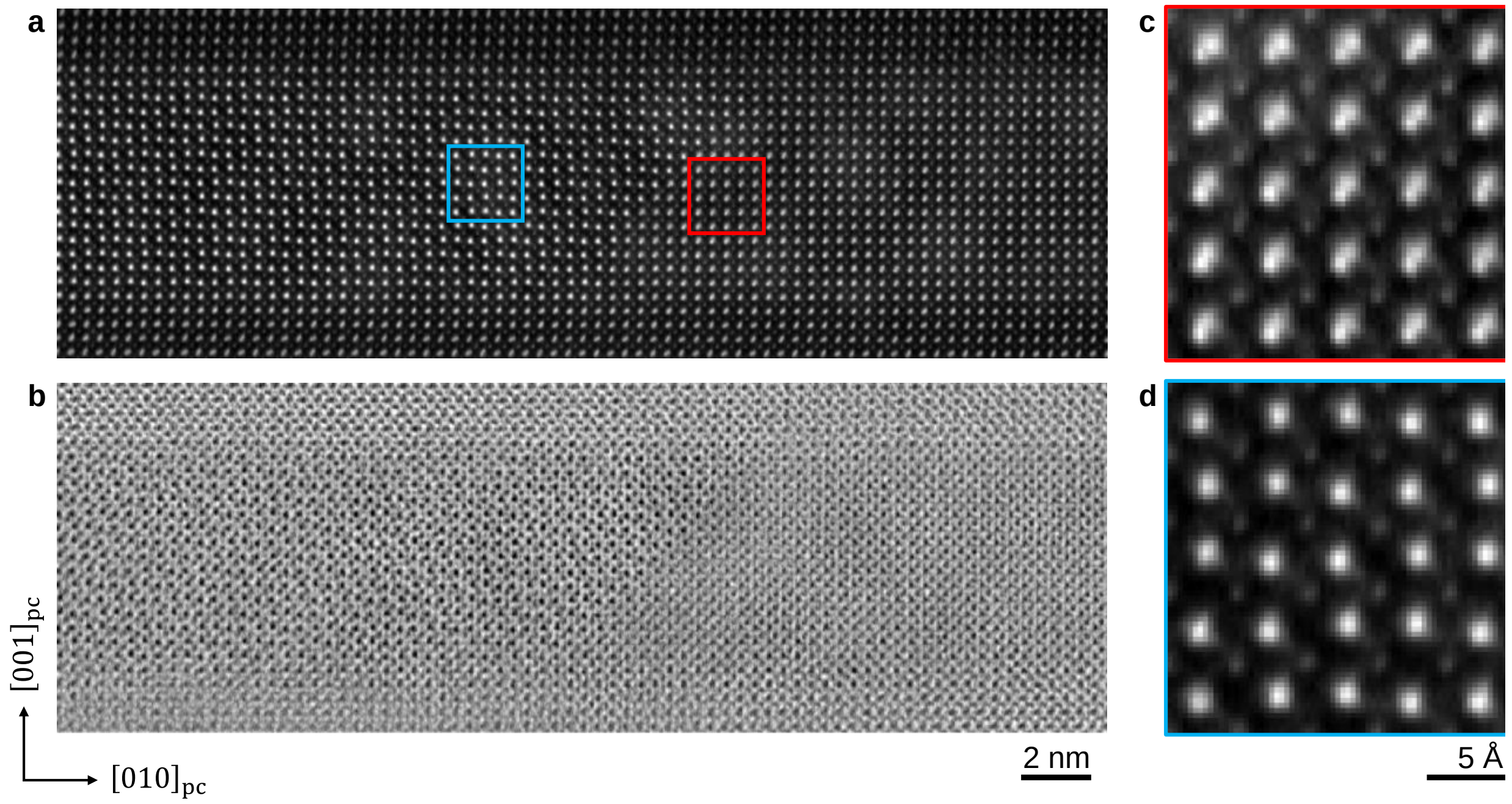

**Figure S6 | Atomic-resolution images of the BFO/TSO superlattice.** Simultaneous **a)** HAADF-STEM and **b)** annular bright-field (ABF-) STEM image of the $[(BFO)_{20}/(TSO)_{10}]_{20}$//TSO superlattice along the $[100]_{pc}$ zone axis corresponding to Fig. 1f in the main text. Drastic contrast differences in ABF-STEM are observed between the antipolar *Pnma*-AFE (left) and polar (right) phases. Zoomed in regions for the **c)** polar and **d)** anti-polar phases from red and blue boxes in (a), respectively. We note the presence of Bi dumbbells in the polar phase, which is not observed in the ground-state polar *R3c* phase of $BiFeO_3$.

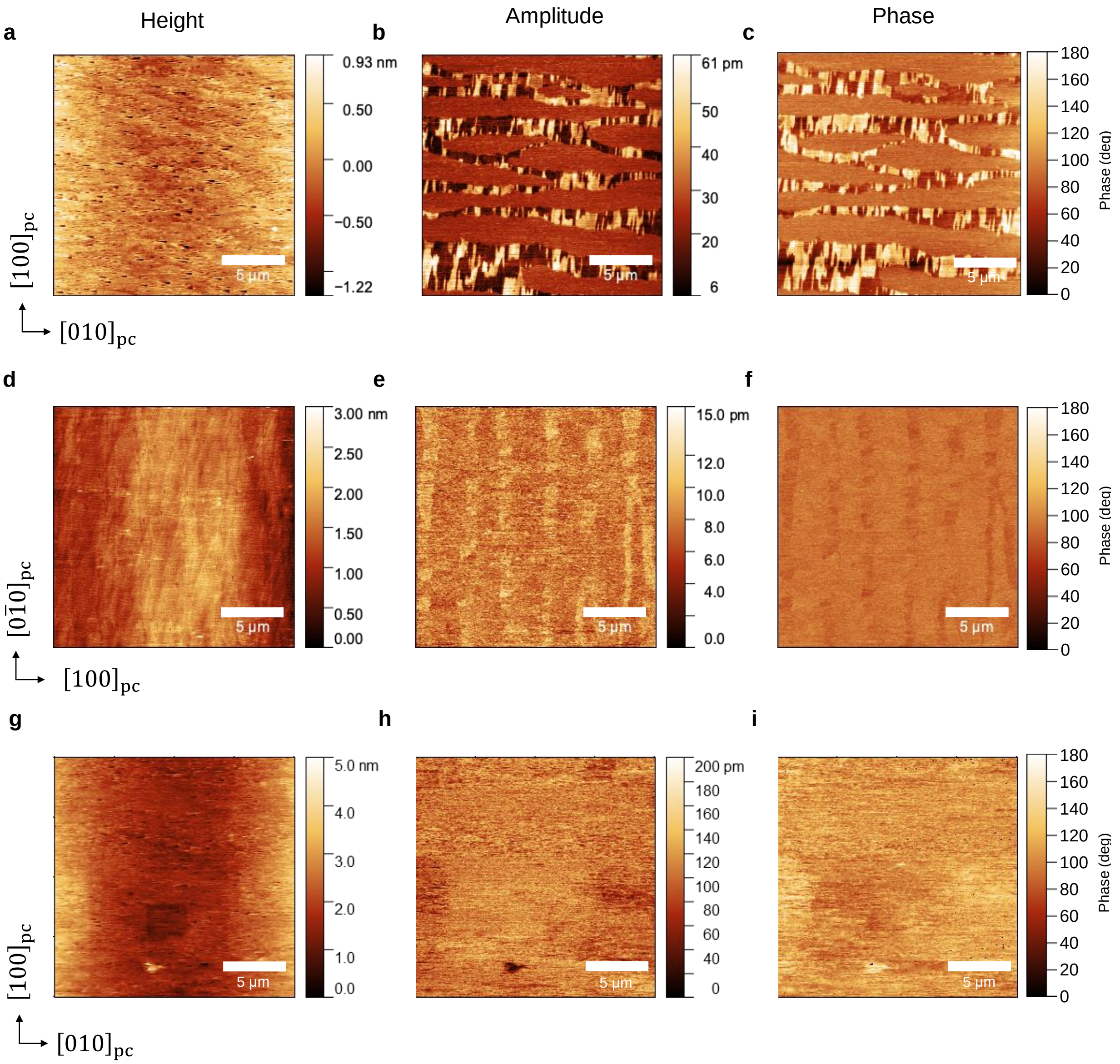

**Figure S7 | Lateral and out-of-plane PFM imaging of mixed-phase coexistence.** Height, amplitude, and phase PFM images for a $[(BFO)_{14}/(TSO)_{10}]_{20}$//GSO superlattice where the amplitude and phase are sensitive to **a-c)** $[100]_{pc}$, **d-f)** $[010]_{pc}$, and **g-h)** $[001]_{pc}$ polar directions. The polar BFO state shows an approximate 180° phase difference along the $[100]_{pc}$ direction, indicating regions of oppositely oriented wave polarization, while the antipolar state shows no response (90° phase). Along the $[010]_{pc}$ direction, minimal signal is observed in the polar state; any residual signal present is likely due to sample misalignment. No signal is observed in the antipolar state (90° phase). Along the out-of-plane $[001]_{pc}$ no signal is present in either state. Scale bars are 5 µm.

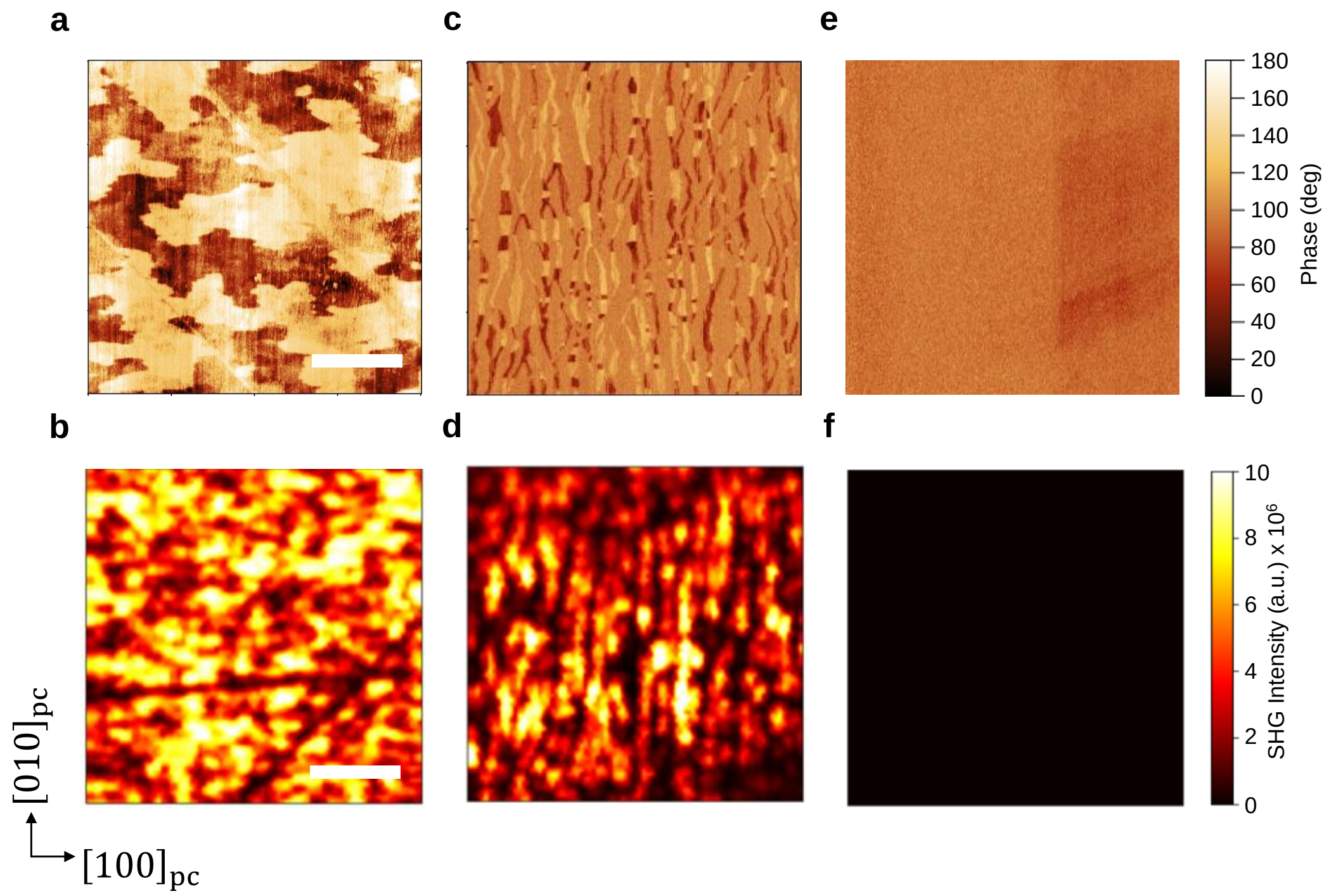

**Figure S8 | BFO thickness dependence of phase co-existence.** $[100]_{pc}$-sensitive PFM phase images and confocal SHG maps for various surface locations on **a,b)** TSO//$[(BFO)_{20}/(TSO)_{10}]_{20}$, **c,d)** TSO//$[(BFO)_{14}/(TSO)_{10}]_{20}$, **e,f)** TSO//$[(BFO)_{11}/(TSO)_{10}]_{20}$ superlattices. a.u., arbitrary units. Scale bars are 5 μm.

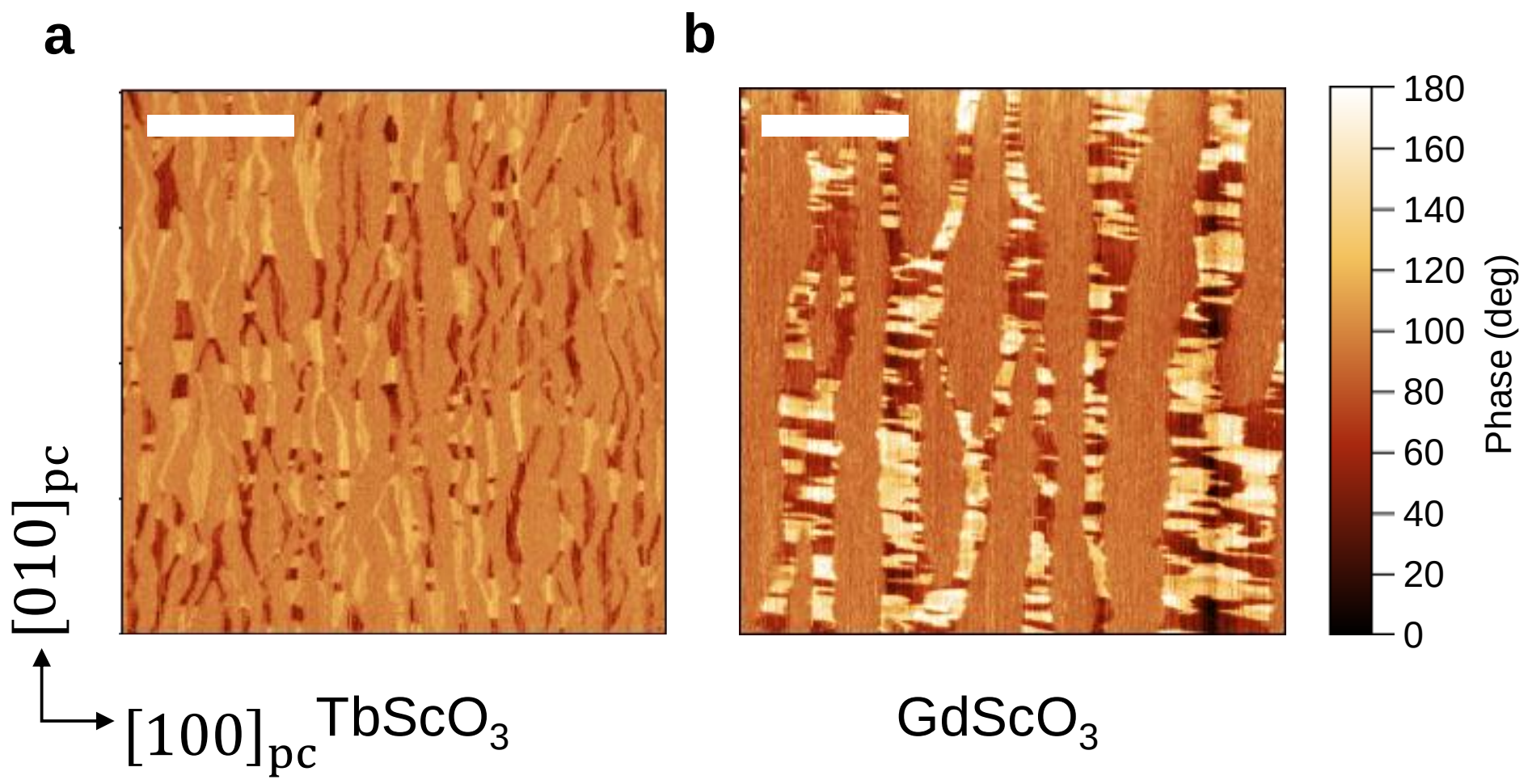

**Figure S9 | Strain dependence of mixed-phase coexistence.** $[(BFO)_{14}/(TSO)_{10}]_{20}$ superlattices grown on **a)** $TbScO_3$ and **b)** $GdScO_3$ substrates, providing tensile and compressive strain on the BFO layers, respectively. Scale bars are 5 μm.

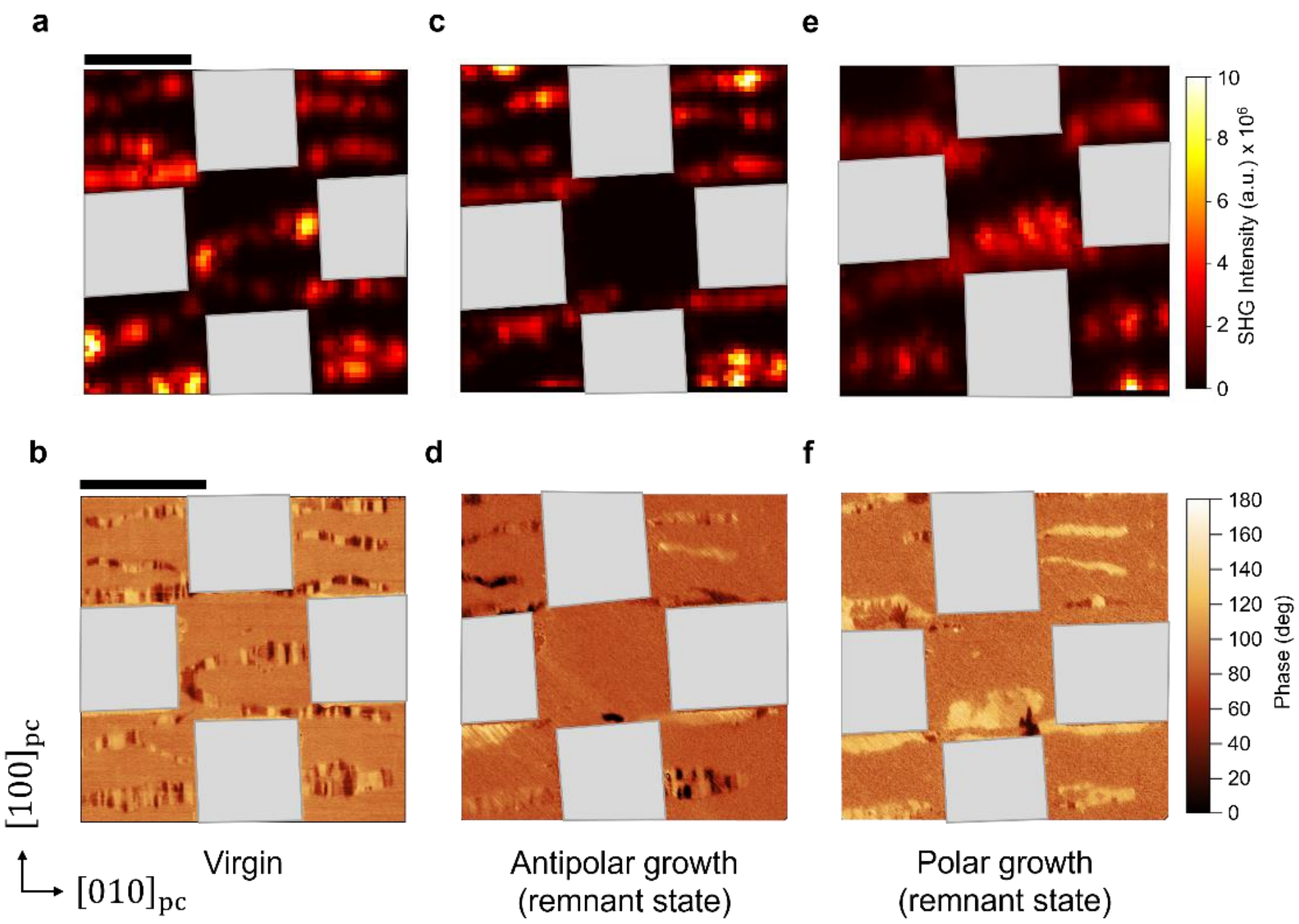

**Figure S10 | Corresponding SHG and PFM imaging with electric field.** Virgin $[(BFO)_{14}/(TSO)_{10}]_{20}$ sample grown on GSO shown with **a)** SHG and **b)** PFM phase imaging on the same area of the sample, where the PFM is sensitive to $[100]_{pc}$ direction of the polarization. Overlaid grey boxes indicate locations of patterned electrodes. **c)** SHG and **d)** PFM phase images of the remnant state of the device shown in (a) and (b) after electric field poling along the $[010]_{pc}$ direction, leaving a near uniform antipolar state. **e)** SHG and **f)** phase images of the remnant state after subsequent poling along the $[100]_{pc}$ direction, regrowing the polar phase domain within the electrodes with the polar orientation along the field direction.

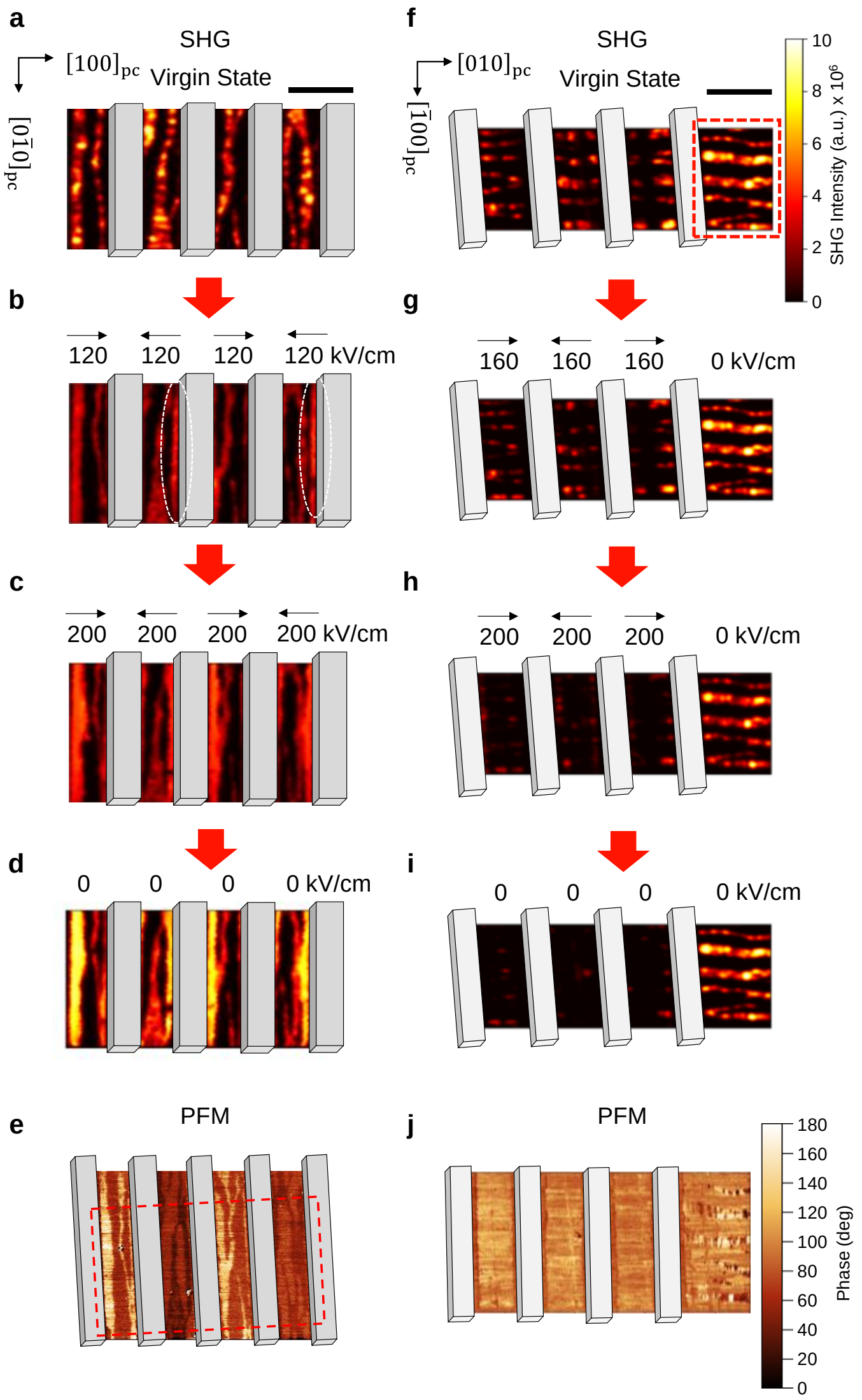

**Figure S11 | SHG and PFM as a function of electric field. a)** Virgin state SHG image of a $[(BFO)_{14}/(TSO)_{10}]_{20}$ superlattice grown on GSO with patterned interdigitated electrodes (IDE), shown in overlaid grey boxes. **b,c)** SHG as a function of lateral electric field along the $[100]_{pc}$ direction on the same area of the device. Magnitude and direction of the electric field is shown above each adjacent region of the IDEs. White circled region indicates location of nucleated polar domains. **d)** Remnant state of the device shown in (a)-(c) probed by SHG after poling to 200 kV/cm. **e)** Remnant state probed by lateral PFM sensitive to the polarization along the $[100]_{pc}$ after poling to 200 kV/cm. Red-outlined region corresponds to the same region as shown in (d). **f)** Virgin state SHG image on the same superlattice with orthogonally-patterned interdigitated electrodes. Red outlined region indicates a section of the sample outside the device area. **g,h)** SHG as a function of lateral electric field along the $[010]_{pc}$ direction on the same area of the device. Magnitude and direction of the electric field is shown above each adjacent region of the IDEs. **i)** Remnant state of the device shown in (g)-(h) measured by SHG after poling to 200 kV/cm. **j)** Remnant state measured by lateral PFM sensitive to the polarization along the $[100]_{pc}$ after poling to 200 kV/cm for the same device region shown in (f)-(i). Scale bar for the entire figure is 10 μm.

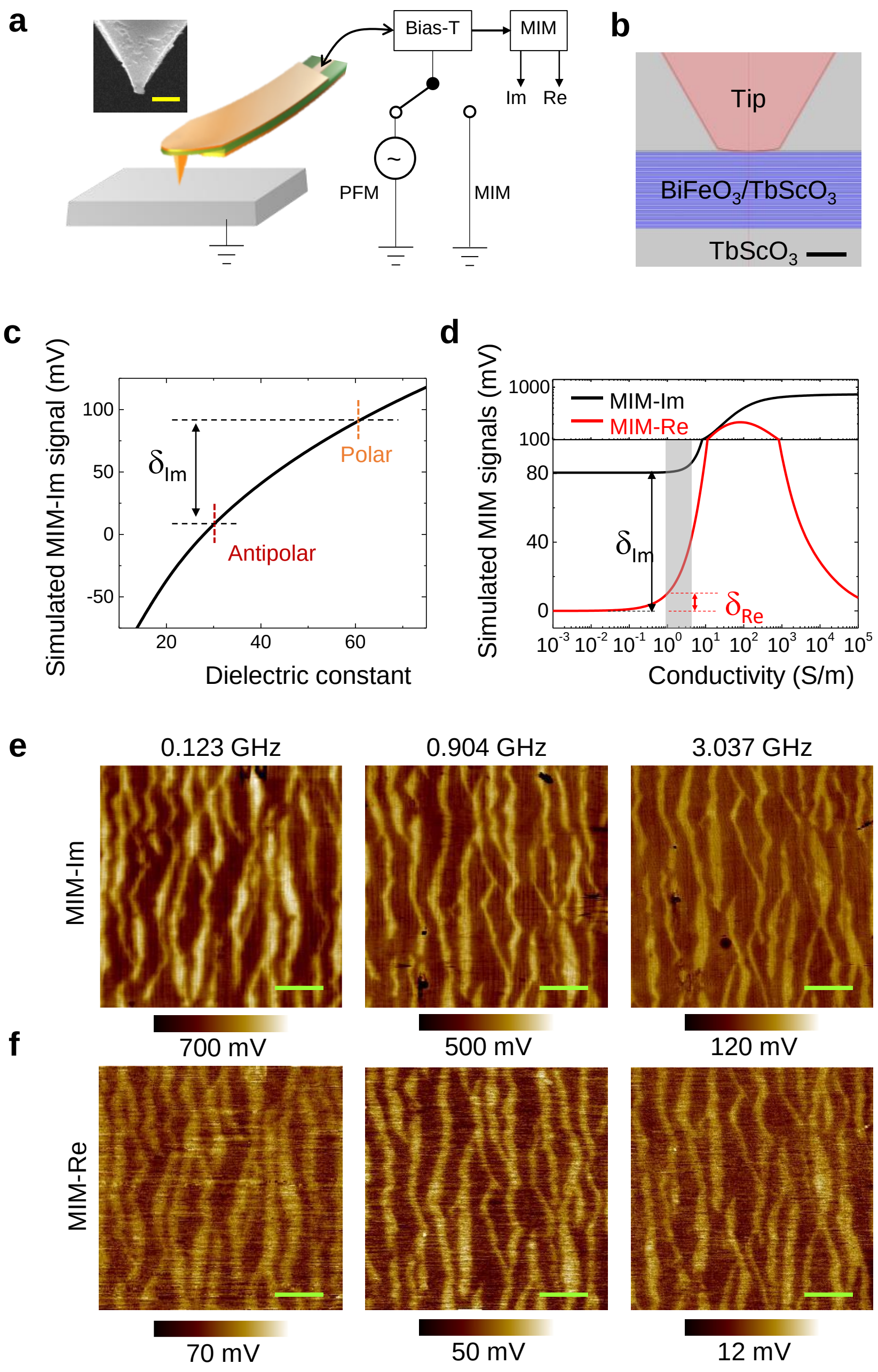

**Figure S12 | MIM data analysis and frequency dependence. a)** Schematic of the PFM / MIM setup. The inset shows a scanning electron micrograph (SEM) near the apex of the tip. Scale bar: 500 nm. **b)** Tip-sample configuration for the finite-element modeling (FEA). Scale bar: 100 nm. **c)** FEA-simulated MIM-Im signal with respect to the TSO substrate ($\varepsilon_r$ = 28) as a function of the dielectric constant of the BFO layer. The permittivity estimated from the measured contrast between the two phases $\Delta_{Im}$ ~ 80 mV is indicated in the plot. **d)** Simulated MIM-Im and MIM-Re signals as a function of the conductivity of the polar phase. The measured contrasts between the two phases in both channels are denoted in the plot. The shaded region indicates the range of conductivity within the measurement error. **e)** Left to right: MIM-Im images at 123 MHz, 904 MHz, and 3.037 GHz. **f)** Left to right: Corresponding MIM-Re images at three illustrative frequencies. Imaging was performed on a $[(BFO)_{14}/(TSO)_{10}]_{20}$//TSO sample. Scale bars: 2 μm.

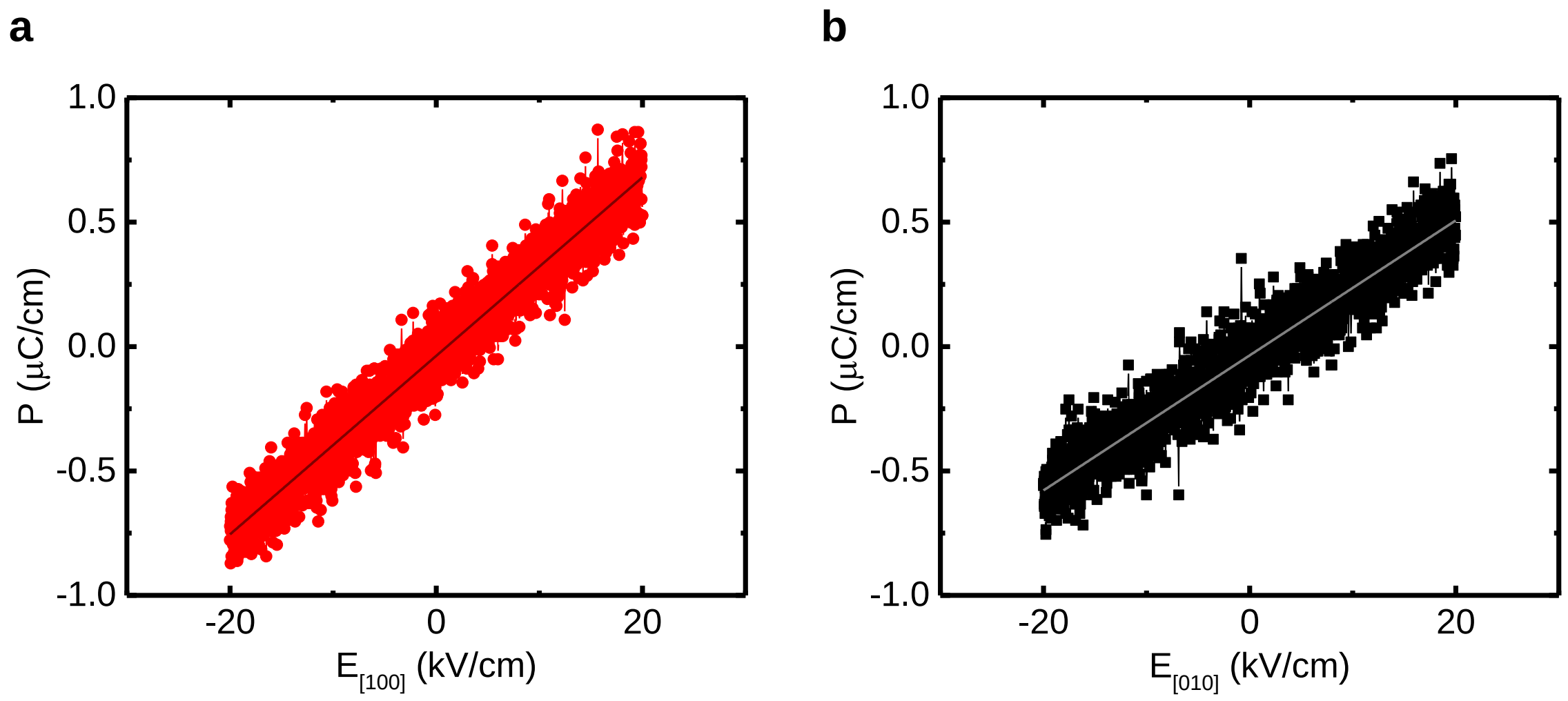

**Figure S13 | Estimation of dielectric constants.** Low electric field dielectric response from **a)** polar and **b)** antipolar $BiFeO_3$ phases measured with interdigitated electrodes on TSO//[(BFO)$_{19}$/(TSO)$_{10}$]$_{20}$ and TSO//[(BFO)$_{11}$/(TSO)$_{10}$]$_{20}$, respectively.

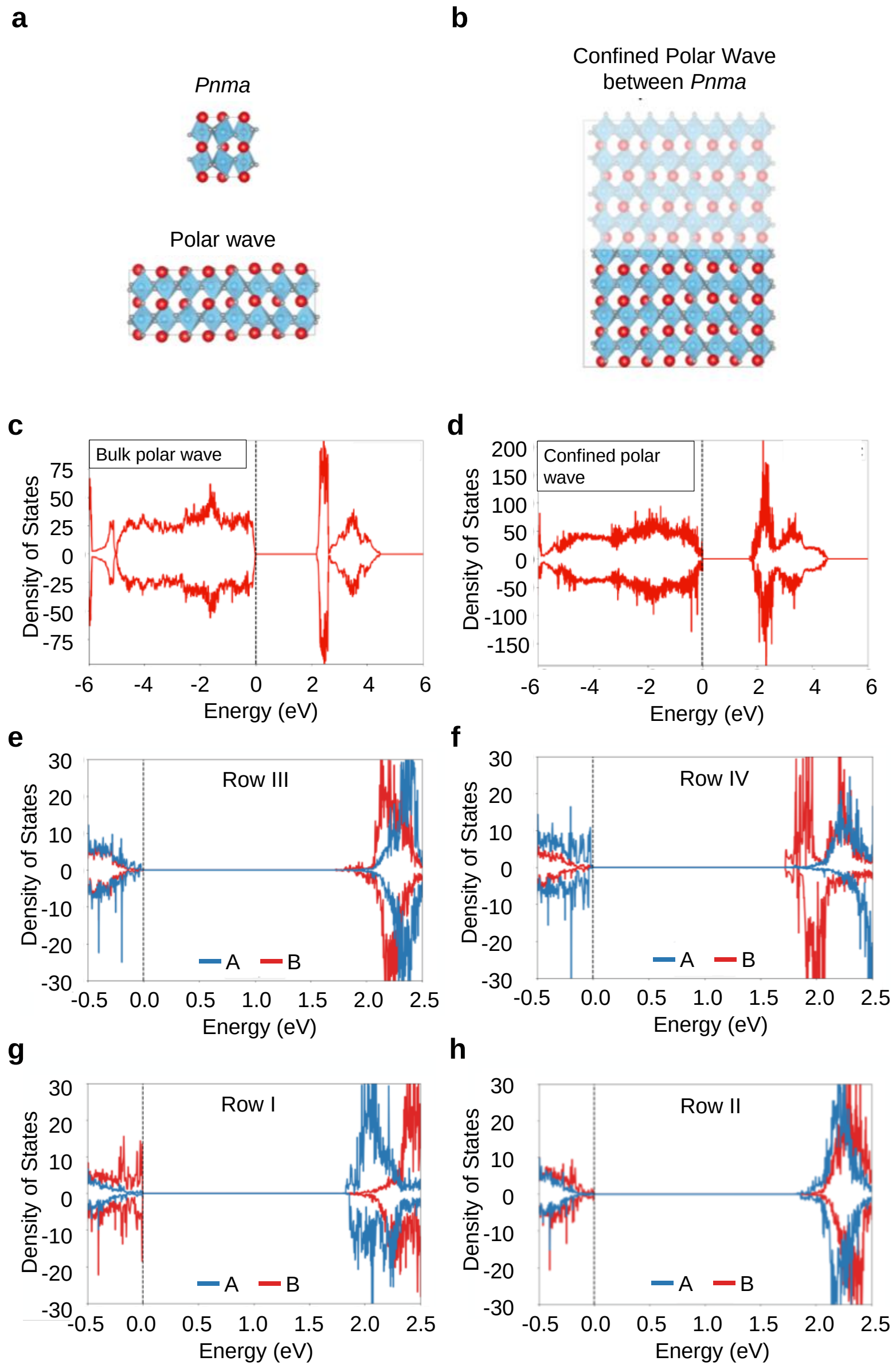

**Figure S14 | Electrostatic model of polarization wave conduction. a)** Atomic models of the bulk *Pnma*-AFE and the *Pc* polarization wave phase of $BiFeO_3$. **b)** Atomic model of the *Pc* polarization wave phase confined by the *Pnma*-AFE phase. DFT-computed density of states for **c)** the isolated, bulk *Pc* wave phase and **d)** the confined *Pc* wave phase shown in (a) and (b), respectively. A reduction in the band gap of ~0.45 eV is observed in the confined polarization wave. **e-h)** DFT-computed density of states for a confined polarization wave heterostructure in the two regions indicated in (d), where a built-in voltage is established by $[001]_{pc}$-oriented component of the wave polarization

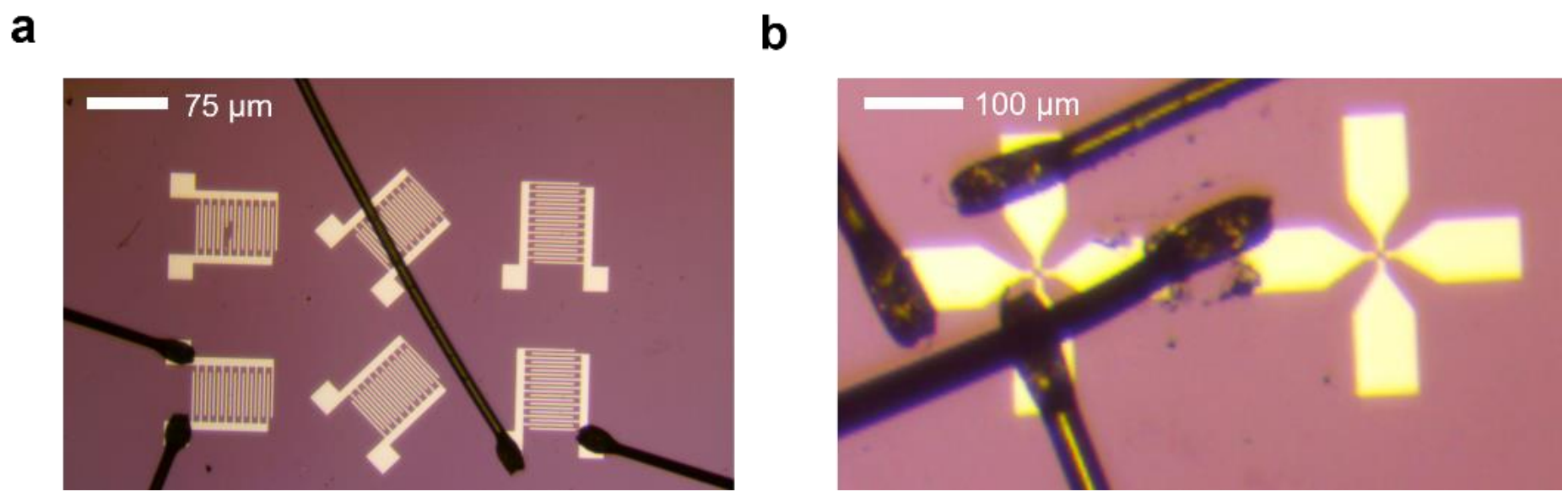

**Figure S15 | Lithographically-defined devices for electric field manipulation. a)** Interdigitated electrodes (IDEs) patterned on the surface of a superlattice sample. IDE spacing ranges from 1 to 10 μm. **b)** Lithographically-defined four-pad devices for orthogonal electric field application. Separation between opposite pads is 6 μm. Wires are gold-bonded to the sample.

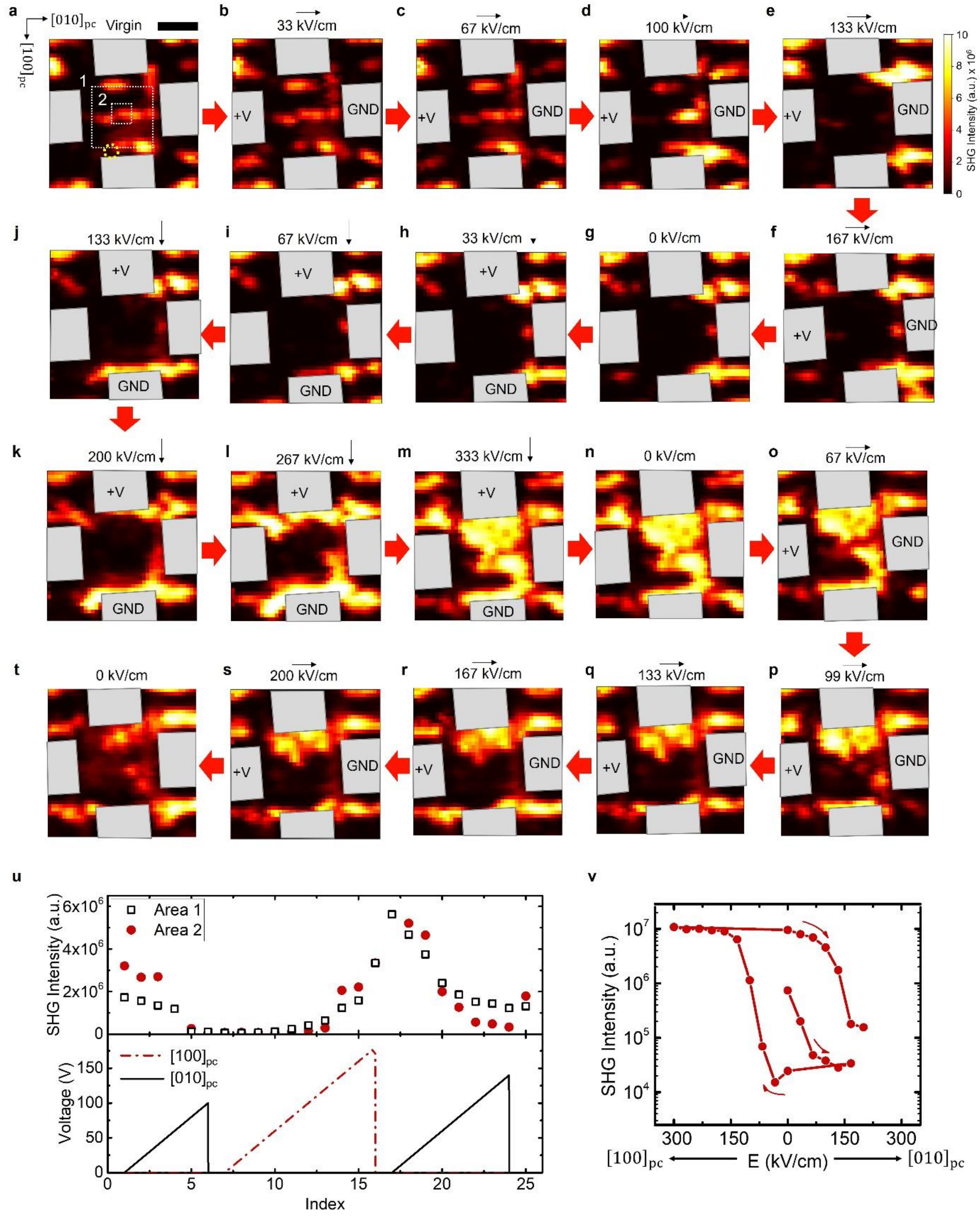

**Figure S16 | *In-situ* SHG as a function of electric field for two orthogonal electric field directions.** SHG images of a GSO//$[(BFO)_{14}/(TSO)_{10}]_{20}$ superlattice with lithographically-patterned 4-point electrodes (shown in grey boxes). **a-f)** A sequence of *in-situ* SHG images in which increasing electric fields are applied along the $[010]_{pc}$. There the magnitude and direction of the electric field is denoted above each panel. The remnant state is shown in **(g)**. **h-m)** Subsequent electric fields are applied along

the [110]$_{pc}$ direction, with the remnant state shown in **(n)**. **o-s)** Finally, starting from the state in **(n)**, an electric field is reapplied along the [010]$_{pc}$ direction, with the remnant state shown in **(t)**. **u)** SHG intensity and applied voltage as a function of SHG image for index for the two regions outline in panel (a). Here panel (a)=1, (b)=1, (c)=3, etc. **v)** An SHG-electric field hysteresis loop for the yellow region circled in (a) showing an approximately three orders of magnitude change in SHG intensity. The positive (negative) horizontal axis denotes electric field oriented along the [010]$_{pc}$ ([100]$_{pc}$) direction.

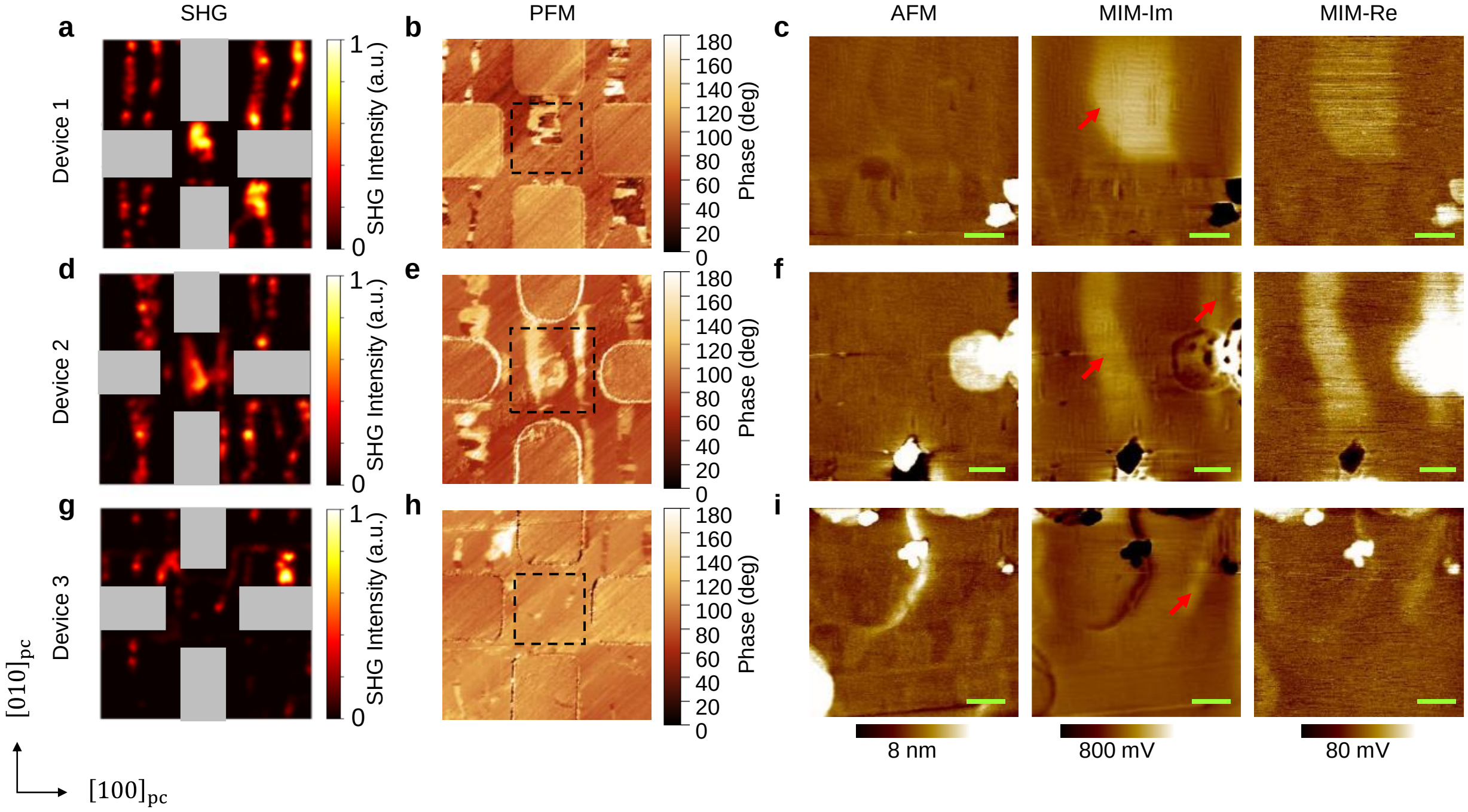

**Figure S17 | Correlated SHG, PFM, and MIM images on poled devices.** SHG, PFM phase, and MIM images (left to right: AFM, MIM-Im, and MIM-Re) on the same location of three poled devices on a $[(BFO)_{14}/(TSO)_{10}]_{20}$//GSO superlattice. The PFM is sensitive to polarization along the [100]$_{pc,}$ and the MIM data is imaged within in the dashed black box in the corresponding PFM image. The red arrows indicate the polar phase seen in the PFM data. **a-c**) shows a device in its virgin state (Device 1). **d-f)** show a device after poling along the [100]$_{pc}$ direction, expanding the polar, non-centrosymmetric phase. **g-i)** show a device after poling along the [010]$_{pc}$ direction, expanding the antipolar, centrosymmetric phase. SHG scale bars: 3 μm. PFM scale bars: 3 μm. MIM scale bars: 1 μm

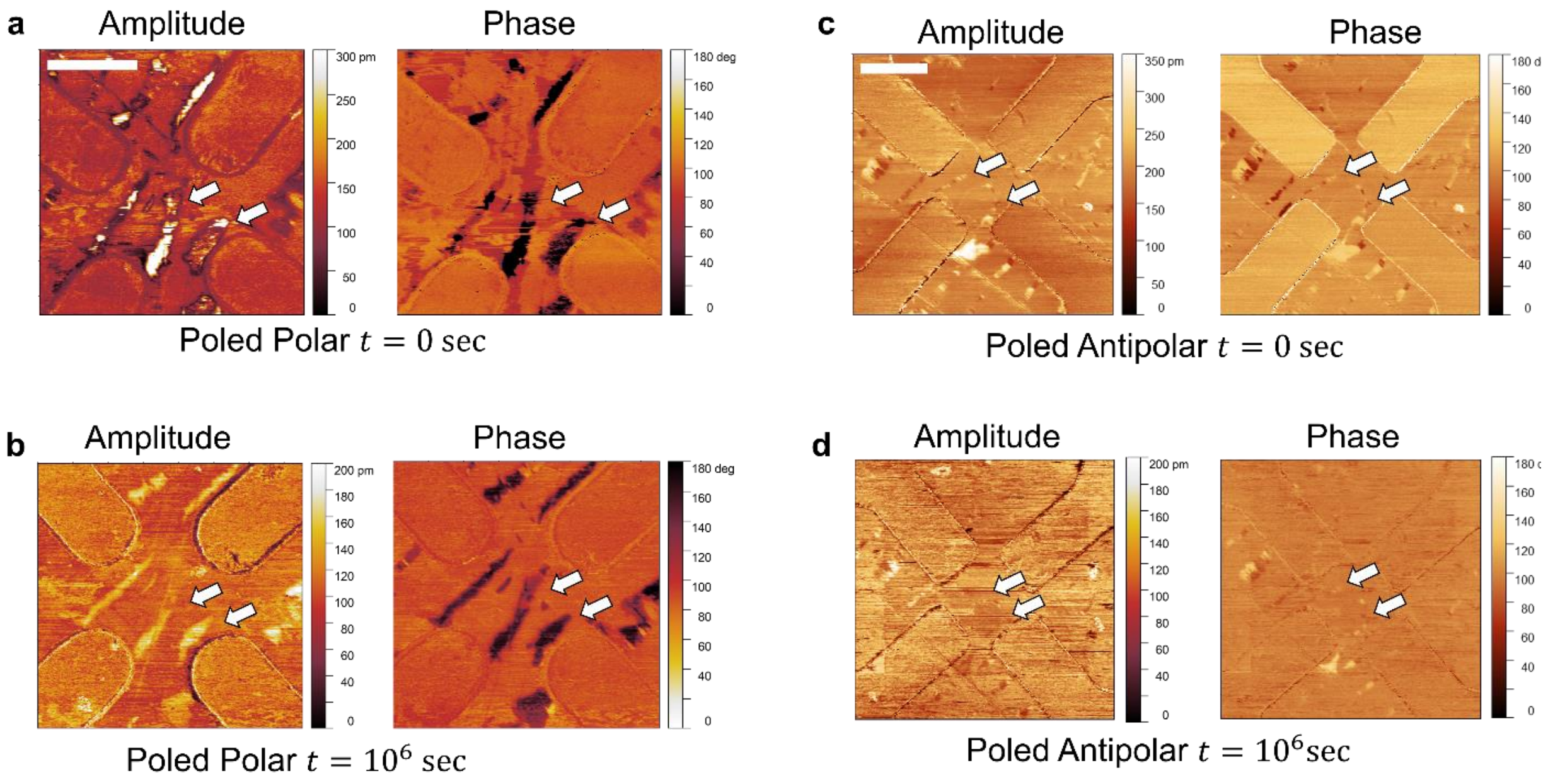

**Figure S18 | Retention of poled states.** Retention (nonvolatility) of polar and antipolar states after nucleation, as shown by amplitude and phase piezoforce response microscope (PFM) images on mixed phase $[(BFO)_{14}/(TSO)_{10}]_{20}$//GSO superlattice samples. **a)** Remanent state of nucleated polar phase domains (indicated by white arrows) immediately after electric field poling of 300 kV/cm. **b)** Retained polar phase domains (indicated by white arrows) >$10^6$ seconds after electric field poling. Nearly full retention is observed. **c)** Remanent state of nucleated antipolar phase domains immediately after electric field poling of 250 kV/cm. White arrows indicate remaining string-shaped domains after poling used as a reference. **d)** Retained antipolar phase domains >$10^6$ seconds after electric field poling. String-shaped domains are noted with white arrows. Nearly full retention is observed. Scale bar = 5 μm

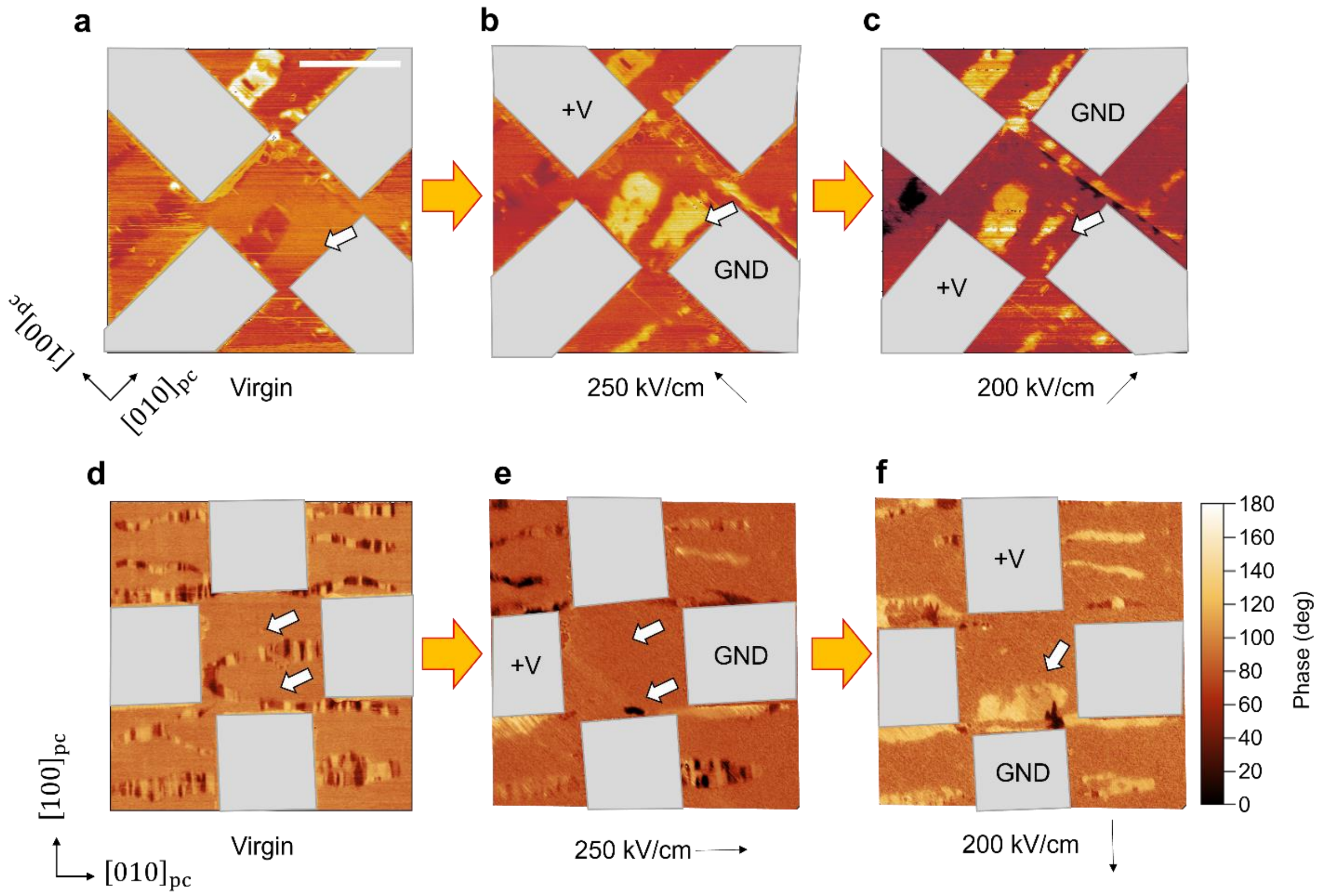

**Figure S19 | Repeatability of poling.** Remanent piezoforce response microscopy (PFM) phase images of two poled devices (a-c) and (d-f), illustrating repeatability of the electric-field-induced phase transformation on a $[(BFO)_{14}/(TSO)_{10}]_{20}$//GSO superlattice. **a)** and **d)** show the virgin state of each device before poling. White arrows indicate regions of interest. **b)** and **e)** show nucleation and growth of polar and antipolar states, respectively, and **c)** and **f)** demonstrate growth of antipolar and polar phase, respectfully.

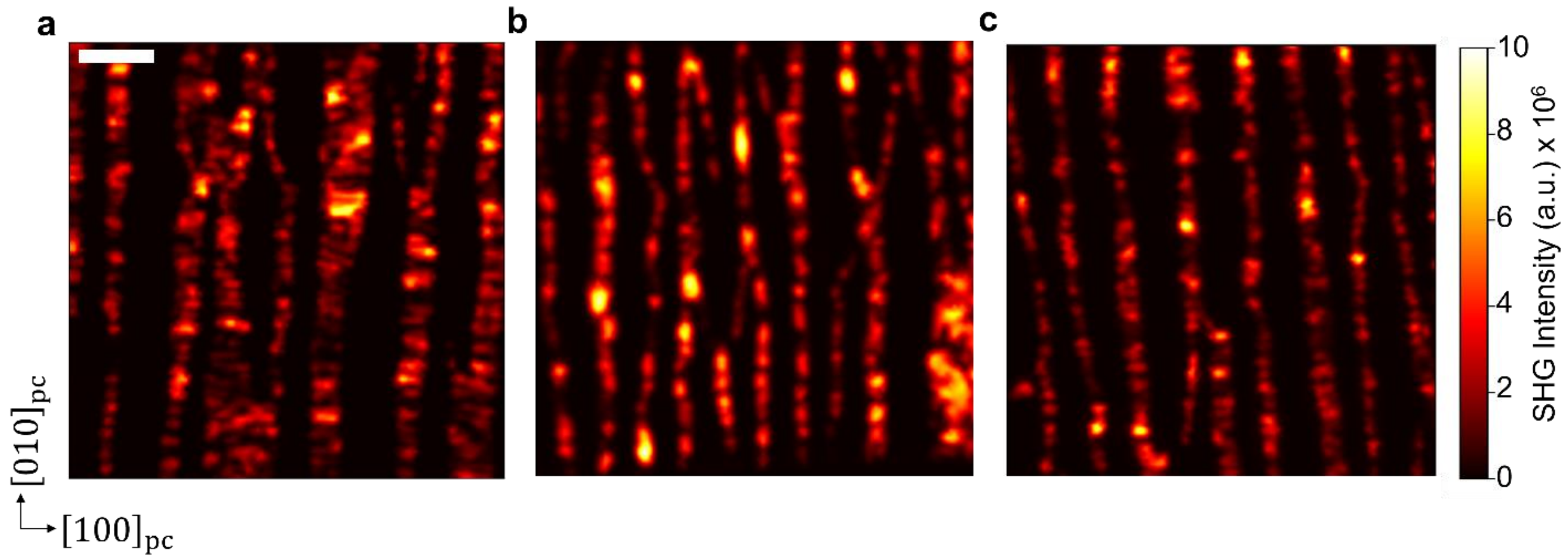

**Figure S20 | Reproducibility of mixed-phase coexistence.** Second harmonic generated (SHG) images from three (a,b,c) nominally-identical $[(BFO)_{14}/(TSO)_{10}]_{20}$//GSO superlattice samples, illustrating reproducibility of the mixed-phase coexistence. Scale bar is 5 µm.

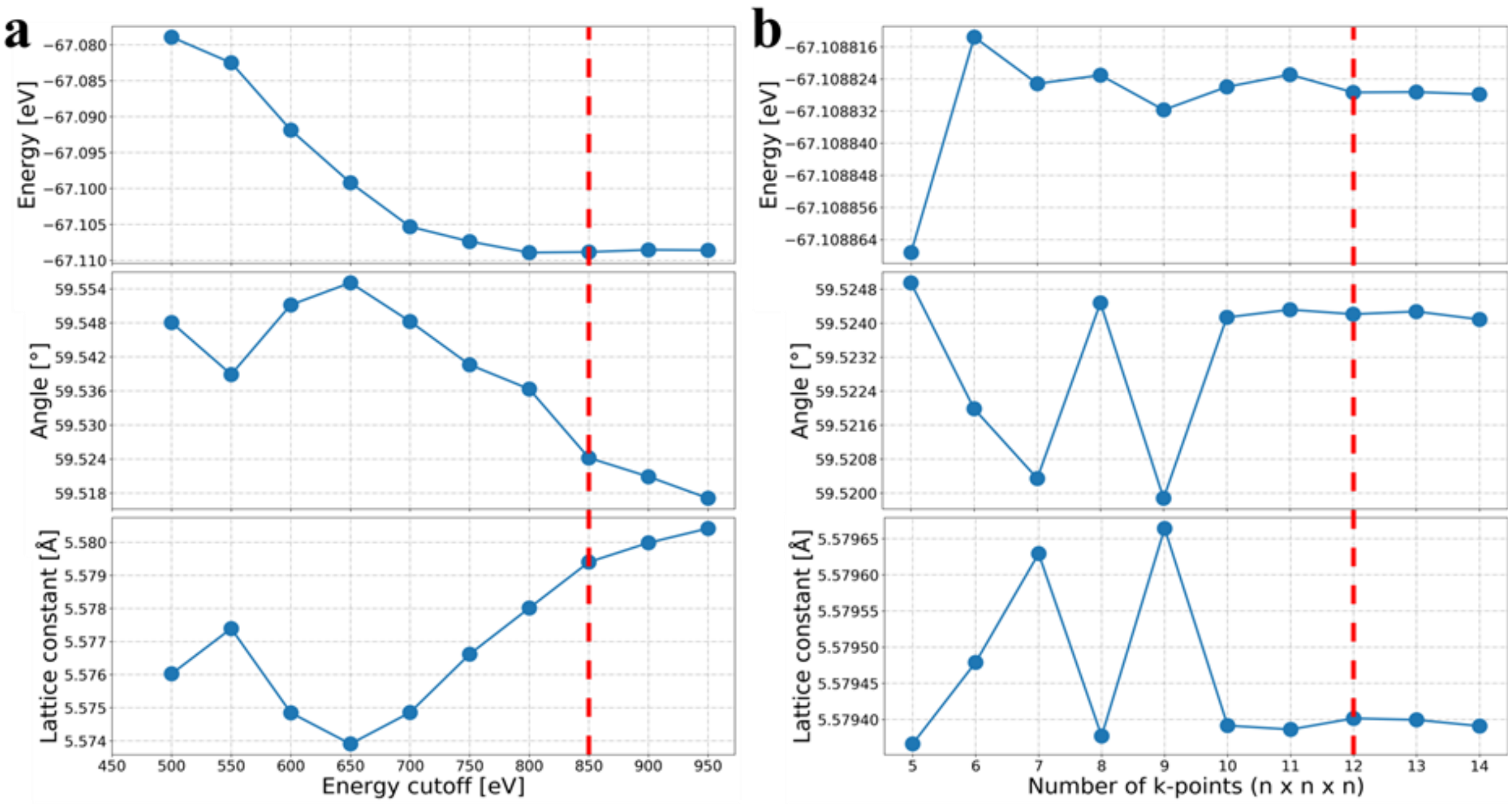

**Figure S21 | DFT convergence tests.** Convergence tests of the numerical parameters performed on the *R*3*c* phase of $BiFeO_3$. From top-to-bottom, the energy, angle and lattice constant of the rhombohedral unit cell as a function of **a)** the energy cutoff and as a function of **b)** the number of k points. The vertical dashed line shows the values that we used in the current work. For the set of parameters that we used, we obtain a lattice constant of 5.5794 Å with a rhombohedral angle of 59.5242°. For reference we can compare these values with the experimental values from [2], where the lattice constant was reported to be 5.63 Å and the angle 59.35°.

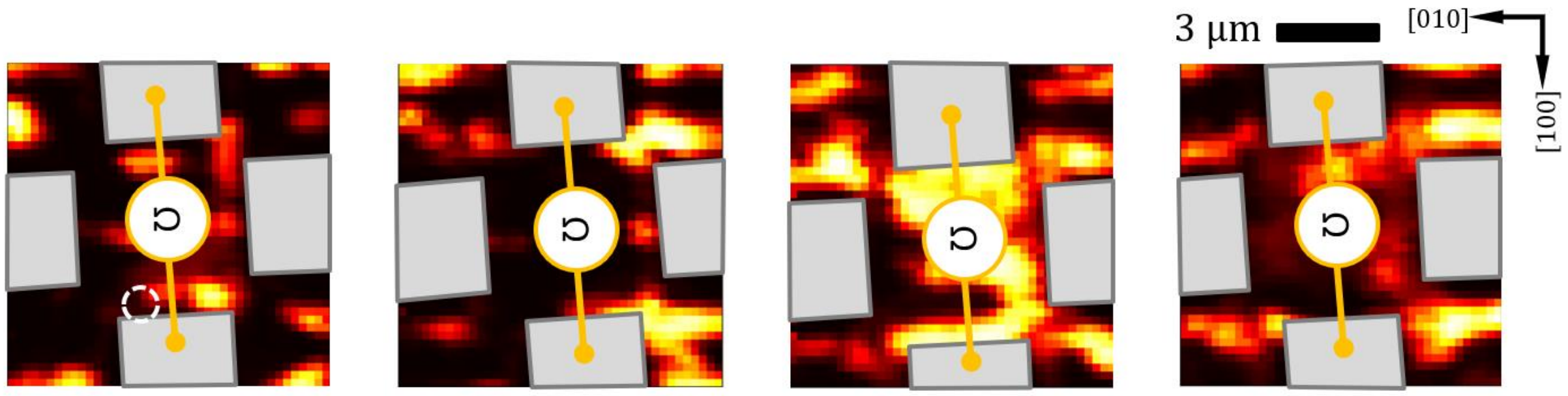

**Figure S22 | Geometry of the *in-situ* resistivity measurement.** The lateral resistivity (ρ) is estimated from the two-point resistance (R) measurement shown above via ρ=R*L/A, where L=6μm is the distance between two electrodes and A is the cross-sectional area of the sample where the electric field is applied. We approximate that the electric field is uniform throughout the thickness of the sample. The resistance is always measured across the electrodes along the $[100]_{pc}$. However, we note that the results are similar if the resistance is measured across the electrodes along the $[010]_{pc}$. In other words, the resistivity is isotropic

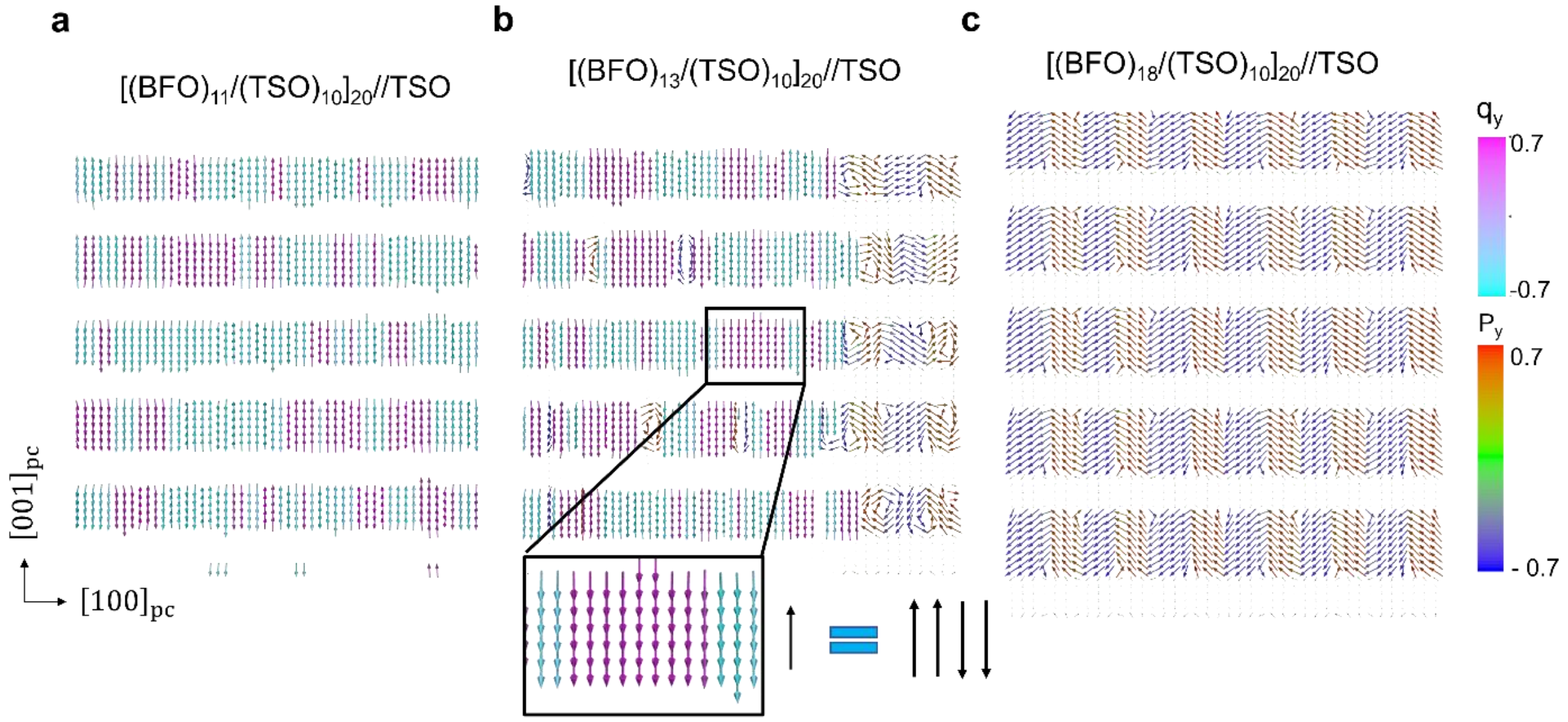

**Figure S23 | Thickness-dependent phase field simulations.** Thickness dependence of the mixed-phase coexistence of polar and antipolar phase in $[(BFO)_n/(TSO)_{10}]_{20}$//TSO superlattices, where **a)** n=11, **b)** n=13, and **c)** n=18 unit cells. These simulations confirm the experimental results (see Extended Data Fig. 8) that at large thicknesses of BFO (>18 unit cells), the BFO is essentially in the polar phase. For very small thicknesses (<11 unit cells), the BFO is essentially in the antipolar phase. For intermediate thicknesses, we observe the mixed-phase coexistence. $q_y$ indicated the direction of the antipolar order parameter and $P_y$ indicates the y-component of the polarization in the polarization wave phase.

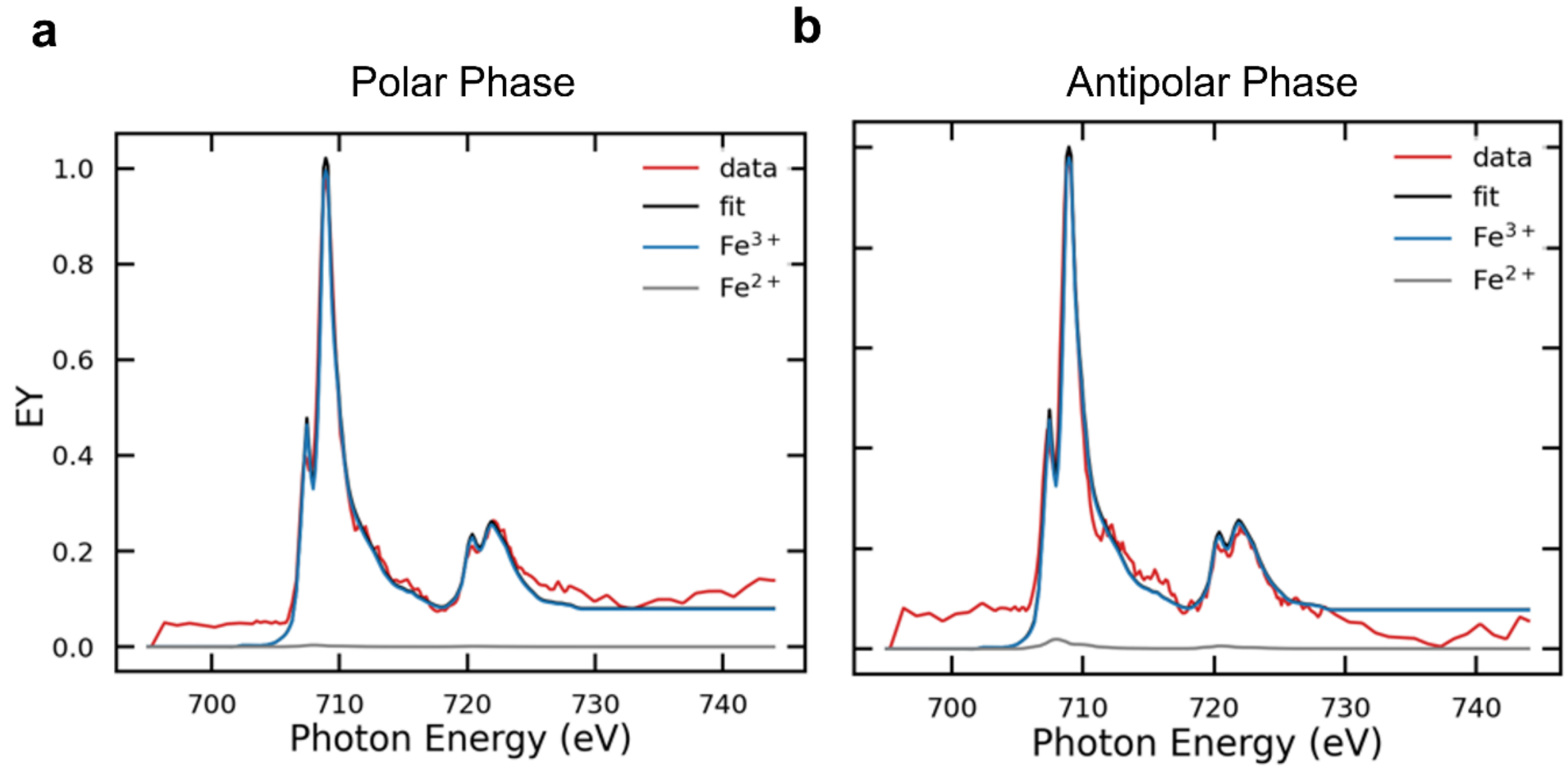

**Figure S24 | X-ray absorption spectra (XAS) of polar and antipolar phases.** Spatially-resolved XAS obtained on both the **a)** polar and **b)** antipolar phases in the $[BFO_{14}/TSO_{10}]_{20}$//GSO sample obtained via photoemission electron microscopy (PEEM) via electron yield mode (EY). The spectra are fit using a linear combination of $Fe^{3+}$ and $Fe^{2+}$ reference spectra from ref [3]. Based on the spectra and the fits, the Fe is almost entirely (>98%) in the 3+ state in both polar and antipolar phases.

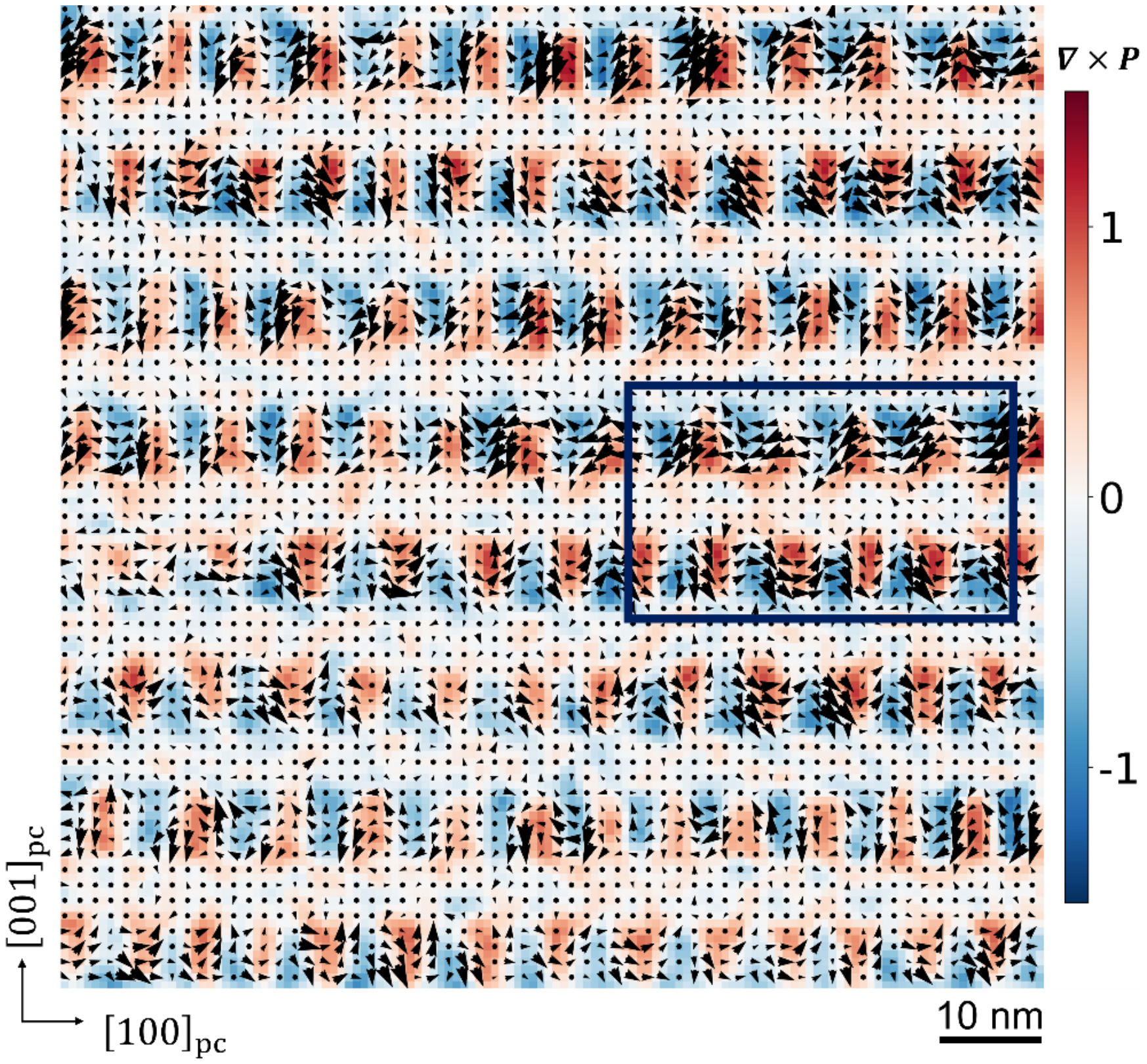

**Figure S25 | Polarization wave phase in the $[(BFO)_{20}/(TSO)_{14}]_{20}$//GSO superlattice.** Polarization map of the polar *Pc* phase overlaid with its curl ($\nabla\times\vec{P}$) obtained by analyzing the Kikuchi bands recorded in the SCBED dataset using an EMPAD. The box indicates the map corresponding to Main Text Fig. 1d. We note that the curl is nonzero for most of the regions within the BFO, indicating the rotation of the polar vector being continuous throughout the phase. This is different from the previously reported 109° domain wall, which the polar vectors only rotates at the domain boundary.

**Table S1 | Description of each superlattice structure and substrate for each Main Text and Extended Data figure.** Superlattice Structure subscripts after parentheses denote unit cells, while subscripts after brackets denote repeats.

| *Manuscript Figure* | *Superlattice Structure* | *Superlattice Substrate* |
|---|---|---|
| 1a,b,c,d,e,f | $[(BFO)_{14}/(TSO)_{10}]_{20}$ | $GdScO_3$ |
| 1g | $[(BFO)_{20}/(TSO)_{10}]_{20}$ | $TbScO_3$ |
| 2 | $[(BFO)_{14}/(TSO)_{10}]_{20}$ | $GdScO_3$ |
| 3 | $[(BFO)_{14}/(TSO)_{10}]_{20}$ | $TbScO_3$ |
| 4 | $[(BFO)_{14}/(TSO)_{10}]_{20}$ | $GdScO_3$ |
| S2 | $[(BFO)_{14}/(TSO)_{10}]_{20}$ | $GdScO_3$ |
| S3 | $[(BFO)_{20}/(TSO)_{10}]_{20}$,<br>$[(BFO)_{14}/(TSO)_{10}]_{20}$,<br>$[(BFO)_{11}/(TSO)_{10}]_{20}$ | $TbScO_3$ |
| S4 | $[(BFO)_{14}/(TSO)_{10}]_{20}$ | $GdScO_3$ |
| S5 | $[(BFO)_{20}/(TSO)_{10}]_{20}$,<br>$[(BFO)_{11}/(TSO)_{10}]_{20}$ | $TbScO_3$ |
| S6 | $[(BFO)_{14}/(TSO)_{10}]_{20}$ | $TbScO_3$ |
| S7 | $[(BFO)_{14}/(TSO)_{10}]_{20}$ | $GdScO_3$ |
| S8 | $[(BFO)_{20}/(TSO)_{10}]_{20}$,<br>$[(BFO)_{14}/(TSO)_{10}]_{20}$,<br>$[(BFO)_{11}/(TSO)_{10}]_{20}$ | $TbScO_3$ |
| S9a | $[(BFO)_{14}/(TSO)_{10}]_{20}$ | $TbScO_3$ |
| S9b | $[(BFO)_{14}/(TSO)_{10}]_{20}$ | $GdScO_3$ |
| S10 | $[(BFO)_{14}/(TSO)_{10}]_{20}$ | $GdScO_3$ |
| S11 | $[(BFO)_{14}/(TSO)_{10}]_{20}$ | $GdScO_3$ |
| S12 | $[(BFO)_{14}/(TSO)_{10}]_{20}$ | $TbScO_3$ |
| S13 | $[(BFO)_{20}/(TSO)_{10}]_{20}$, $[(BFO)_{11}/(TSO)_{10}]_{20}$ | $TbScO_3$ |
| S16 | $[(BFO)_{14}/(TSO)_{10}]_{20}$ | $GdScO_3$ |
| S17 | $[(BFO)_{14}/(TSO)_{10}]_{20}$ | $GdScO_3$ |

**Supplementary Text**

1. *CBED symmetry determination*

CBED is a well-established and powerful technique for determining crystal point and space-group symmetry at the nanoscale[4–6]. CBED patterns are obtained by focusing a convergent electron probe on a local region of the specimen, where in modern electron microscopes the electron probe sizes can range from sub-nm to several microns. Due to the probe convergence, the Bragg reflections appear as disks instead of sharp spots. As a result of dynamical diffraction effects, the crystal point group can thus be determined by a careful analysis of intensity patterns, also known as rocking curves, within the transmitted and diffracted disks.

For determining the (absence) presence of inversion symmetry in (non)centrosymmetric crystals, we carefully selected experimental CBED patterns with the highest symmetry and compared with simulated patterns of a known space group. For example, the antipolar phase in $BiFeO_3$ layer has a centrosymmetric space group of *Pnma* which exhibits *2mm* point group symmetry along the $[1\bar{1}0]_O$ zone axis. This *2mm* symmetry is reflected in both experimental (Fig. S4f) and simulated (Fig. S4g) CBED patterns, with the mirror planes *m* along directions of $m \parallel [110]_O$ and $m \parallel [001]_O$. In the case of polar crystals, the polarization leads to the breakdown of Friedel's due to dynamical diffraction effects, which results as intensity asymmetry of Friedel pairs in the mirror plane along certain projections (white arrows, Fig. S4k).

While zero-order Laue zone (ZOLZ) reflections in CBED patterns are useful for determining mirror and rotational symmetry elements a projection, the high-order Laue zone (HOLZ) detail reveals the symmetry elements present in the three-dimensional unit cell. Specifically, the HOLZ ring radius is inversely proportional to the square-root of the lattice-spacing along the electron beam projection direction. Using $TbScO_3$ along the $[1\bar{1}0]_O$ zone axis as a reference, where the HOLZ ring in Fig. S4d corresponds to a periodicity of ~7.9 Å, the antipolar phase in $BiFeO_3$ layer (Fig. S4h) exhibits a HOLZ ring with $\sim\frac{1}{\sqrt{2}}\times$ in radius corresponding to a periodicity of ~15.8 Å, which is consistent with ~4× large of a primitive perovskite unit cell.

2. *Charge-voltage hysteresis loops of polar and antipolar phases*

Figures S5a and b show hysteresis loops for the polar and antipolar phases, respectively. Figure S5a was taken on $[(BFO)_{20}/(TSO)_{10}]_{20}$, which contains only the polar state with the electric field oriented along the $[100]_{pc}$ direction. The polar hysteresis loop is nearly fully remnant with a polarization of $\sim 10\ \mu C/cm^2$. Figure S5b shows a PE loop taken on $[(BFO)_{11}/(TSO)_m]_{20}$ which stabilizes only the antipolar state. The electric field was applied along the $[010]_{pc}$. Characteristic double switching, typical of antiferroelectric materials, is present. This antipolar state has a saturation polarization of $\sim 10\ \mu C/cm^2$, and a remnant polarization near zero, consistent with results observed elsewhere[7].

Panels (d) and (e) in Fig. S5 show PFM phase images of the polar remnant states of $[(BFO)_{20}/(TSO)_{10}]_{20}$, as indicated in panel (a) on the same area of the interdigitated electrode. The PFM phase measurement is sensitive to polarization along the $[100]_{pc}$. The area to the left of the electrodes is outside of the active area of the device and serves as a reference and alignment region. Clearly visible is the 180° phase shift in the PFM following the application of antiparallel electric fields along the $[100]_{pc}$ direction, visually confirming the switching behavior present in panel (a). We note a partially switched area in panel (e) in the red dashed oval.

*3. Thickness- and strain-dependence of emergent phases*

To further elucidate the origins of the phase coexistence of centrosymmetric (antipolar) and non-centrosymmetric (polar) $BiFeO_3$ (BFO) phases, superlattices were grown with varying thicknesses of BFO and various strain states governed by the underlying substrate choice. The relative stability of each phase can be controlled by engineering the thickness of the BFO layer, i.e., tuning electrostatic and interface contributions. Figure S8 shows lateral piezoforce microscopy (PFM) phase images for BFO/TSO superlattices grown on (110) oriented TSO substrates, which contain BFO thicknesses of 19 unit cells (Fig. S8a), 14 unit cells (Fig. S8b), and 11 unit cells (Fig. S8c) respectively. The contrast in the PFM phase images is sensitive to polarization along the $[100]_{pc}$ direction. Notably, when the BFO thickness is large (19 unit cells), only the polar *Pc* phase is stabilized, as shown by the high response cow-like pattern domains in Fig. S8a. Light (Dark) regions are polar domains oriented along the $[100]_{pc}$ ($[\bar{1}00]_{pc}$) direction. Conversely, samples containing relatively thin BFO (11 unit cells) show no PFM response, and thus only stabilize the antipolar *Pnma*-AFE phase of BFO. For a BFO thickness of 14 unit cells, a mixture of both polar and antipolar phases co-exists forming stripe-like patterns, where within the polar regions 180$^o$ domain boundaries separate oppositely-oriented polar regions. The remarkably dramatic evolution of BFO phases over an eight unit-cell variation in the BFO thickness indicated a highly phase tunability. Moreover, it suggests a first-order phase transition between polar and anti-polar phases and a potential energetic pathway between the two phases using external stimuli.

Choice of sample substrate can change the mechanical strain state of the epitaxially-grown film. In Fig. S9, we compare the PFM phase response of nominal $[(BFO)_{14}/(TSO)_{10}]_{20}$ films grown on $GdScO_3$ (GSO) and TSO substrates. The GSO substrate places BFO under a lateral tensile strain, while the TSO substrate places BFO under a lateral compressive state. Surprisingly, both phases are still present regardless of whether the BFO is under tensile or compressive strain, at least for the strains provided by these substrate choices. In fact, the relative amounts of each phase also stay essentially the same, suggesting that the stability is robust against strain variations. However, although the amount of each phase remains the same in the two of strain states (approximately 50:50 ratio of each phase), the length scales of the stripe domains changes by as much as a factor of 10.

Although the thickness- and substrate-dependent stability of the polar state versus the antipolar phase is not the central focus of this paper, it is noteworthy that such small changes in the thickness of the individual layers can have profound impact on the stability of the two states.

*4. Microwave Impedance Microscopy Analysis and Modeling*

The PFM and MIM images shown in Fig. 3 of the main text were taken by the same shielded cantilever probe[8]. As illustrated in Fig. S12a, we first performed the PFM measurement by applying kHz voltages on the tip through a bias-tee, and then grounded the tip for MIM experiments at various microwave frequencies. We verified the mixed-phase coexistence by performing PFM and MIM ($f$ = 0.123 GHz) imaging on the same locations of the poled devices, as shown in Fig. 3 of the Main Text and Fig. S17. The one-to-one correlation between the two sets of data on three different devices clearly demonstrates the robustness of our conclusions. For data analyses, we modeled the near-field interaction by using the actual tip and sample (20 periods of BFO and TSO layers) configurations shown in Fig. S12b. Here the tip shape is based on the SEM image near the apex, as seen in the inset of Fig. S12a. Using a calibration sample[9], we obtained a conversion coefficient of ~ 0.6 mV/nS from the simulated admittance (in unit of nS) to the measured signals (in unit of mV), which is needed to extract the permittivity and conductivity values in Fig. 3. Finally, we took similar MIM images at frequencies ranging from ~ 100 MHz to 3 GHz, as displayed in Fig. S12 (c) and (d). Since these images were acquired by different microwave electronics, the absolute signal levels cannot be directly compared. Using the same calibration process as above, we confirmed that the dielectric constants of the two phases are independent of frequency within our measurement range. Similarly, the conductivity of the polar phase is around 0.1 ~ 1 S/m from the MIM-Re signals.

*5. Estimation of low frequency dielectric constants*

The dielectric constants of the polar and antipolar phases were measured on the pure polar and antipolar phase superlattice samples, $TSO//[(BFO)_{19}/(TSO)_{10}]_{20}$ and $TSO//[(BFO)_{11}/(TSO)_{10}]_{20}$, respectively (Fig. S13), using in-plane interdigitated electrodes (IDEs, see Fig. S15). The low electric-field responses for each phase are shown in Fig. S14, where the slope of the curve yields the permittivity. Each curve was measured at 1 kHz. The permittivity of each phase can be estimated by approximating the device as a parallel plate capacitor, $C = \frac{KA}{t} n = \frac{dQ}{dV}$, where $K$ is the permittivity of the material, $A$ is the cross sectional area of the device, and $t$ is the thickness of the device. In this geometry, $A = \text{sample thickness} * \text{IDE arm length}$ and $t = \text{IDE arm separation}$. $n$ refers to the number of capacitors in parallel (i.e.- the number of IDE arms). $dQ/dV$ is the change in charge with voltage, related to the slope of the curves in Fig. S13. Using this approximation, the relative dielectric constants of the polar and antipolar phases are estimated to be 62 and 35, respectively.

*6. Second harmonic generation as a function of electric field for orthogonally in-plane electric fields*

Here, we measure the SHG response of the film under in-plane electric fields using two device geometries. Figure S11 show a series of SHG and PFM images of the mixed phase $[(BFO)_{14}/(TSO)_{10}]_{20}$ superlattice with an interdigitated electrode (IDE) micropatterned on the surface of the film (also see Methods and Fig. S15). Polar phases are denoted by high SHG intensity while antipolar regions show no SHG response. The IDE allows the application of in-plane electric fields to the sample. The Ta(4 nm)/Pt(50 nm) IDE electrodes are schematically denoted by three-dimensional overlaid grey boxes. Figs. S11a-d show a series of SHG images under progressively increasing applied lateral electric fields along the $[100]_{pc}$ and $[\bar{1}00]_{pc}$ directions. Electric fields in between adjacent electrodes are oriented antiparallel, as denoted by the arrows above the images. Electric fields were applied for approximately two minutes between image acquisitions. Under relatively low electric fields, the SHG signal becomes more uniform and increases in intensity within the polar (non-centrosymmetric) regions of the sample. Further increases in the electric field cause nucleation of the polar phase from the antipolar phase as seen in the white circled regions on the device in Fig. S11b, and an enhancement of the SHG contrast. Subsequent growth of the nucleated and already existing domains occurs on further increasing the field (Fig. S11b-d), providing further evidence of the first-order transition between polar and antipolar phases. Figure S11d and S11e show the remnant state of the superlattice after poling using both SHG (Fig. S11d) and PFM (Fig.

S11e). A clear correlation in the size and shape of each domains is present between the two techniques, verifying the phase transformation observations. As expected, the in-plane orientation of the polar phase measured by PFM lies along the direction of the in-plane electric field applied between the IDEs, where regions between adjacent electrodes experience opposite electric fields and contain oppositely polarized states.

A similar image sequence is shown in Fig. S11f-j, where the electric field is instead applied along the orthogonal $[010]_{pc}$ and $[0\bar{1}0]_{pc}$ directions on a different region of the sample. In this image sequence, the area of the film denoted with a red dashed outline remains unpoled and always in its virgin state, whereas the areas of the film between the electrodes are poled. Here, regardless of a positive or negative applied field (seen in adjacent electrodes), the SHG signal reduces with increasing magnitude of electric field, implying a transformation of the non-centrosymmetric polar state into the centrosymmetric antipolar state. Domain expansion of the antipolar phase can be seen with increasing electric field, in agreement with the first-order phase transition hypothesis. It is noteworthy that the unpoled region remains relatively unimpacted and provides a reference state of SHG contrast. Figure S11i and S11j show the remnant state of the superlattice measured using SHG (Fig. S11i) and PFM (Fig. S11j). A clear correlation is observed between the SHG and PFM, where polar regions are denoted by high response regions in each imaging technique, whereas antipolar regions are denoted by low response regions. Within the electrodes, little response is seen, whereas the unpoled regions still contain a mixture of polar and antipolar phases.

*7. Interconversion between non-centrosymmetric and centrosymmetric phases in fine electric field steps*

Figure S16 shows a sequence of SHG images under an applied electric field from a lithographically-patterned four-point gate device (see Methods and Fig. S15). The device design allows orthogonal electric fields to be successively applied to the same area of the sample. Figures S16a-g show in-situ SHG images as a function of electric field oriented along the $[010]_{pc}$ direction. The virgin mixed-phase coexistence transforms to the near-uniform centrosymmetric antipolar phase, as indicated by the black contrast (or lack of SHG signal) in the active region of the device between the electrodes in Fig. S16f. In this transformation, we observe a shrinkage of the polar domains consistent with a first-order transition consisting of nucleation and growth, in which the electric field drives the change in position of the boundary between the two phases. Figure S16g shows the remnant state and illustrates the non-volatility of the process. The reversibility of this phase transformation is demonstrated in Fig. S16h-m, where a voltage is applied along the orthogonal $[100]_{pc}$, regrowing the non-centrosymmetric polar phase by driving the phase boundary back. The nonvolatile nature is captured in Fig. S16n. Finally, the antipolar phase can be re-established by reapplying an electric field along the $[010]_{pc}$ direction (Fig. S16o-t). Figure S16u plots the integrated SHG intensity and voltage as a function of index (image in the sequence) for two square regions defined in Fig. S16a). The "colossal" manipulation of nonlinear optical response (SHG) is best illustrated in Fig. S16v, which plots the SHG intensity as a function of electric field, with the positive horizontal axis indicates electric field oriented along the $[010]_{pc}$ direction, and the negative horizontal axis indicating electric field oriented along the orthogonal $[100]_{pc}$ direction.

We note that the effects of fringing fields on the domain state are observed in the image sequence in Fig. S16a-p. In regions outside the active area of the device, changes in the domain structure are observed, which are dependent on the direction of electric field applied. Quantitative data is only taken within the electrode region, where the electric field is well-defined.

*8. Correlative in-situ second harmonic generation (SHG) and piezoforce microscopy (PFM) imaging*

Although a direct comparison between PFM and SHG intensities is nontrivial, SHG is sensitive to crystal symmetry and thus the presence of an SHG signal does correlate with the presence of the non-centrosymmetric polar phase in this material system, whose net polarization can be also probed with PFM. Likewise, the absence of SHG signal corresponds to the presence of the centrosymmetric antipolar phase in this system and the absence of a PFM response. Figure S10 compares SHG and PFM on a BFO/TSO superlattice containing a four-point electrode device, where images were taken at remanence on the same device following different stages of electric field poling. The SHG-imaged domains do not directly reproduce the PFM domains because of SHG interference effects between the resolution-limited domains. The *outline* of the polar/antipolar phases is correctly reproduced but not the interior. Figure S10a,b show the virgin state of the device. In both images, the as-grown polar stripe-like phase is seen along the $[010]_{pc}$ direction. The presence or absence of a SHG signal directly correlates to the presence of the polar or antipolar phase, respectively. On the other hand, phase contrast in PFM correlates to the orientation of the polarization. Regions of 90º phase show no PFM response and are, thus, antipolar. Within the high response regions containing the non-centrosymmetric polar phase, a 180º phase difference is seen in the PFM, which correlate to oppositely-oriented domains

Subsequently, an electric field of ~300 kV/cm was applied to the device along the $[010]_{pc}$ direction. The remnant state (Figs. 10c,d) shows no SHG signal and no PFM response within the gated region, which corresponds to the presence of the centrosymmetric antipolar phase. Additionally, the PFM image clearly shows the effects of fringing electric fields from the device on the nearby striped regions outside the device area, where certain stripes become unform in color (PFM phase of either 0ºor 180º), indicating a preference for one orientation of the polar order. The polar orientation in those regions depends on the fringing field direction.

Figures S10e,f show remnant SHG and PFM images following the subsequent application of an electric field along the $[100]_{pc}$ direction. An electric field along the $[100]_{pc}$ direction favors the non-centrosymmetric polar phase and results in the growth of the polar domain in the active region of the device. This is clearly shown by the emergence of SHG intensity in the lower half of the device, as well as the response in the PFM image. It is noteworthy that the nucleated polar phase in the device is nearly uniformly polarized along the direction of the applied electric field, as shown by the uniform contrast in the PFM image (0 degree PFM phase). A similar sequence of correlated images for poled devices is shown in Fig. S17, which includes PFM phase, SHG, and MIM data.

9. *Estimation of effective band gap difference between polar and antipolar regions*

The conductivity of the polar(antipolar) phase can be written as[10],

$$\sigma_{P(AP)} = n_{P(AP)} e \mu_{P(AP)},$$

where $n_{P(AP)}$ is the number density of carriers in the polar(antipolar) phase, $e$ is the electron charge, and $\mu_{P(AP)}$ is the carrier mobility in the polar(antipolar) phase. Assuming $\mu_P \sim \mu_{AP}$ then $\frac{\sigma_P}{\sigma_{AP}} \sim \frac{n_p}{n_{AP}}$. The concentration of carriers in each phase $n_{P(AP)}$ can in turn be approximated as a function of temperature ($T$), band gap ($E_g$), and intrinsic carrier concentration, $\eta_0$ of each phase[10]:

$$n_{P(AP)} = \eta_0^{P(AP)} \exp\left(-\frac{E_g^{P(AP)}}{kT}\right),$$

where $k$ is the Boltzmann constant. If we assume the intrinsic carrier concentration $\eta_0^{P(AP)}$ is the same in each phase, then

$$\frac{\sigma_P}{\sigma_{AP}} \sim \frac{n_P}{n_{AP}} \sim \frac{\exp\left(-\frac{E_g^P}{kT}\right)}{\exp\left(-\frac{E_g^{AP}}{kT}\right)}$$

The experimentally-observed conductivity ratio between each phase $\frac{\sigma_P}{\sigma_{AP}} \sim 10^5$ (see Fig. 4f), therefore the difference in effective bandgap $\Delta E_g$ required to realize this conductivity ratio is,

$$\Delta E_g \sim kT ln(10^5) \approx 0.3 \text{ eV}$$